\documentclass[traditabstract]{aa} 
\usepackage{graphicx}
\usepackage{txfonts}
\usepackage{natbib}

\def\msol{M$_\odot$}
\def\msun{M$_\odot$}
\def\s{$\sigma$}

\begin{document}
   \title{A MAD view of Trumpler 14\thanks{Based on  observations obtained with the MCAO Demonstrator (MAD) at the VLT Melipal Nasmyth focus (ESO public data release). Tables \ref{tab: catal} and \ref{tab: pairs} are only available in electronic form at the CDS via anonymous ftp to cdsarc.u-strasbg.fr (130.79.128.5) or via http://cdsweb.u-strasbg.fr/cgi-bin/qcat?J/A+A/}}

\authorrunning{H. Sana et al.}
   \author{
H. Sana\inst{1,2},
Y. Momany\inst{1,3},
M. Gieles\inst{1},
G. Carraro\inst{1},
Y. Beletsky\inst{1},
V.~D. Ivanov\inst{1},
G. De Silva\inst{4},
G. James\inst{4}
          }

   \institute{European Southern Observatory, Alonso de Cordova 3107, Vitacura, Santiago 19, Chile\\
              \email{hsana@eso.org}
         \and
             Sterrenkundig Instituut 'Anton Pannekoek', Universiteit van Amsterdam, Science Park 904, 1098 XH Amsterdam, The Netherlands\\
         \and
             INAF,  Osservatorio Astronomico di Padova,  Vicolo dell'Osservatorio  5,  I-35122 Padova, Italy \\
         \and
             European Southern Observatory, Karl-Schwarzschild-Str. 2, D-85748 Garching bei M\"unchen, Germany\\
             }

   \date{Received September 15, 1996; accepted March 16, 1997}

  \abstract
{We present adaptive optics (AO) near-infrared observations of the core of the Tr~14 cluster in the Carina region obtained with the ESO multi-conjugate AO demonstrator, MAD. Our campaign yields AO-corrected observations with an image quality of about 0.2\arcsec\ across the 2\arcmin\ field of view, which is the widest AO mosaic ever obtained. We detected almost 2000 sources spanning a dynamic range of 10~mag. The pre-main sequence (PMS) locus in the colour-magnitude diagram is well reproduced by Palla \& Stahler isochrones with an age of 3 to $5\times10^5$~yr, confirming the very young age of the cluster.  We derive a very high (deprojected) central density $n_0\sim4.5(\pm0.5)\times10^4$~pc$^{-3}$ and estimate the total mass of the cluster to be about $\sim4.3^{+3.3}_{-1.5}\times10^3$~\msun, although contamination of the field of view might have a significant impact on the derived mass. We show that the pairing process is largely dominated by chance alignment so that physical pairs are difficult to disentangle from spurious ones based on our single epoch observation. Yet, we identify 150 likely bound pairs, 30\% of these with a separation smaller than 0.5\arcsec\ ($\sim$1300AU). We further show that at the 2$\sigma$\ level massive stars have more companions than lower-mass stars and that those companions are respectively brighter on average, thus more massive. Finally, we find some hints of mass segregation for stars heavier than about 10~\msun. If confirmed, the observed degree of mass segregation could be explained by  dynamical evolution, despite the young age of the cluster. }


   \maketitle
%

\section{Introduction}\label{sect: intro}

Massive stars do not form in isolation. 
They are born and, for most of them, are living in OB 
associations and young clusters \citep{MAWG04}. Indeed,
 most of the field cases are 
runaway objects that can be traced back to their natal
cluster/association \citep{dWTP05}. Even the best cases of field
massive stars are now questioned in favour of an 
ejection scenario \citep{GvB08}.

One of the most striking and important properties of high-mass stars
is their high degree of multiplicity. Yet accurate observational 
constraints of the multiplicity properties and of the underlying
parameter distributions are still lacking. These quantities are however critical 
as they trace the final products of high-mass star formation and
early dynamical evolution. In nearby open clusters, the 
 minimal spectroscopic binary (SB) fraction is in the range of 40\%\ to 60\%\ 
 \citep{SGN08, SGE09, SJG09}, which is similar to the 57\%\ SB fraction observed for the  galactic O-star population as a whole \citep{MHG09}.

While spectroscopy  is suitable to detect the short- and intermediate-period binaries ($P<10$~yr),
adaptive optics (AO) observations can tackle the problem from the
 other side of the separation range \citep[see e.g.\ discussion in][]{SaLB09}. As an example, \citet{TBR08} obtained
a minimal fraction of massive stars with companions of
37\%\ within an angular separation of 0.2 to 6\arcsec. However, their survey is limited to objects
with declination $\delta > -42$\degr. It is thus missing some of the
most interesting star formation regions of the Galaxy, like the Carina nebula
region. 

In this context, we undertook a multi-band NIR AO campaign on the main Carina region clusters with the Multi-Conjugate Adaptive Optics (MCAO) Demonstrator \citep[MAD,
][]{MBD07}. Beside deep NIR photometry of the individual clusters, our 
survey was designed to provide us with high-resolution imaging of 
the close environment of a sample of 60 O/WR massive stars in the Carina region. Unfortunately, 
the bad weather at the end of the second MAD demonstration run in January
2008 prevented the completion of the project. Valuable $H$ and $K_\mathrm{S}$
photometry of the sole Tr~14 cluster could be obtained. The 2\arcmin\
field of view (fov) still provides us with high-quality information of
the surrounding of $\sim$30 early-type stars with masses above 10~\msol. It also
constitutes the most extended AO mosaic ever acquired. \\

\begin{table} 
\centering
\caption{Field centering (F.C.) for on-object (Tr~14) and on-sky observations and coordinates of the natural guide stars (NGSs).}
 \label{tab: fov}
\begin{tabular}{cccc}
\hline
\hline
          & RA         &         DEC & V mag \\
\hline 
Tr~14 F.C.& 10:43:55.00 & $-$59:33:03.0 & \dots \\
Sky F.C. & 10:43:06.53 & $-$59:37:46.1 & \dots \\
\\
NGS1     & 10:43:59.92 & $-$59:32:25.4 &  9.3  \\
NGS2     & 10:43:57.69 & $-$59:33:39.2 & 11.2  \\
NGS3     & 10:43:48.82 & $-$59:33:24.8 & 10.7  \\
\hline
\end{tabular}
\end{table}

\begin{table}
\centering
\caption{Log of the MAD observations of Tr~14. }
 \label{tab: diary}
\begin{tabular}{ccccccc} 
\hline 
\hline
DP & RA & DEC & DIT & NDIT & $N_\mathrm{IMA}$ & Tot.  \\
\hline
\\
\multicolumn{7}{c}{Trumpler 14 observations}  \\ 
\\
\#1 & 10:43:56.0 & $-$59:32:46 &  2s & 30 & 2$\times$14 & 28 min \\
\#2 & 10:43:56.5 & $-$59:33:29 &  2s & 15 & 2$\times$8 & 8 min \\
\#3 & 10:43:53.5 & $-$59:33:28 &  2s & 15 & 2$\times$8 & 8 min \\
\#4 & 10:43:53.5 & $-$59:32:41 &  2s & 15 & 2$\times$8 & 8 min \\
\\
\multicolumn{7}{c}{Sky field observations (MCAO in open loop)}\\
\\
\#1 & 10:43:07.5 & $-$59:37:29 & 2s & 30 & 8 & 8 min \\
\#2 & 10:43:08.0 & $-$59:38:12 & 2s & 15 & 8 & 4 min \\
\#3 & 10:43:05.0 & $-$59:38:11 & 2s & 15 & 8 & 4 min \\
\#4 & 10:43:05.0 & $-$59:37:24 & 2s & 15 & 8 & 4 min \\
\hline
\end{tabular}
\end{table} 

   \begin{figure}
   \centering
   \includegraphics[width=6cm]{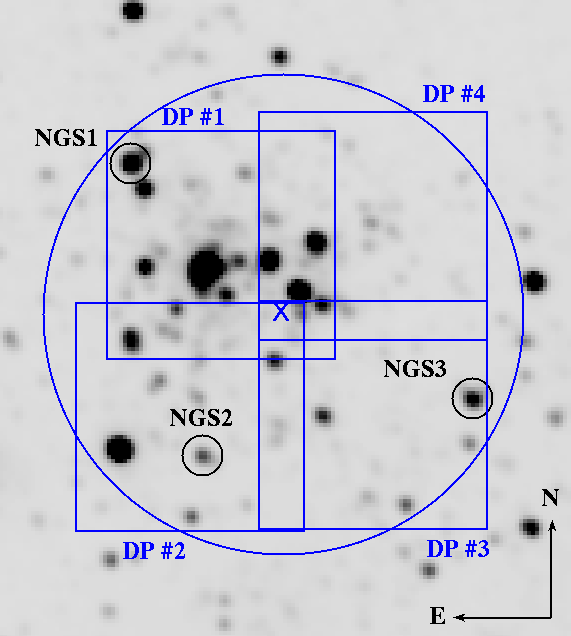}
      \caption{2MASS $K$ band image of Tr~14. The cross and the large 
circle indicate the centre and the size of the 2\arcmin\ MAD field of view.
 The four 57\arcsec$\times$57\arcsec\ boxes show the position 
of the CAMCAO camera in the adopted 4-point dither
pattern, while the selected NGSs are identified by the smaller circles.
              }
   \label{fig: fov}
   \end{figure}

   \begin{figure}
   \centering
   \includegraphics[width=\columnwidth,angle=-90]{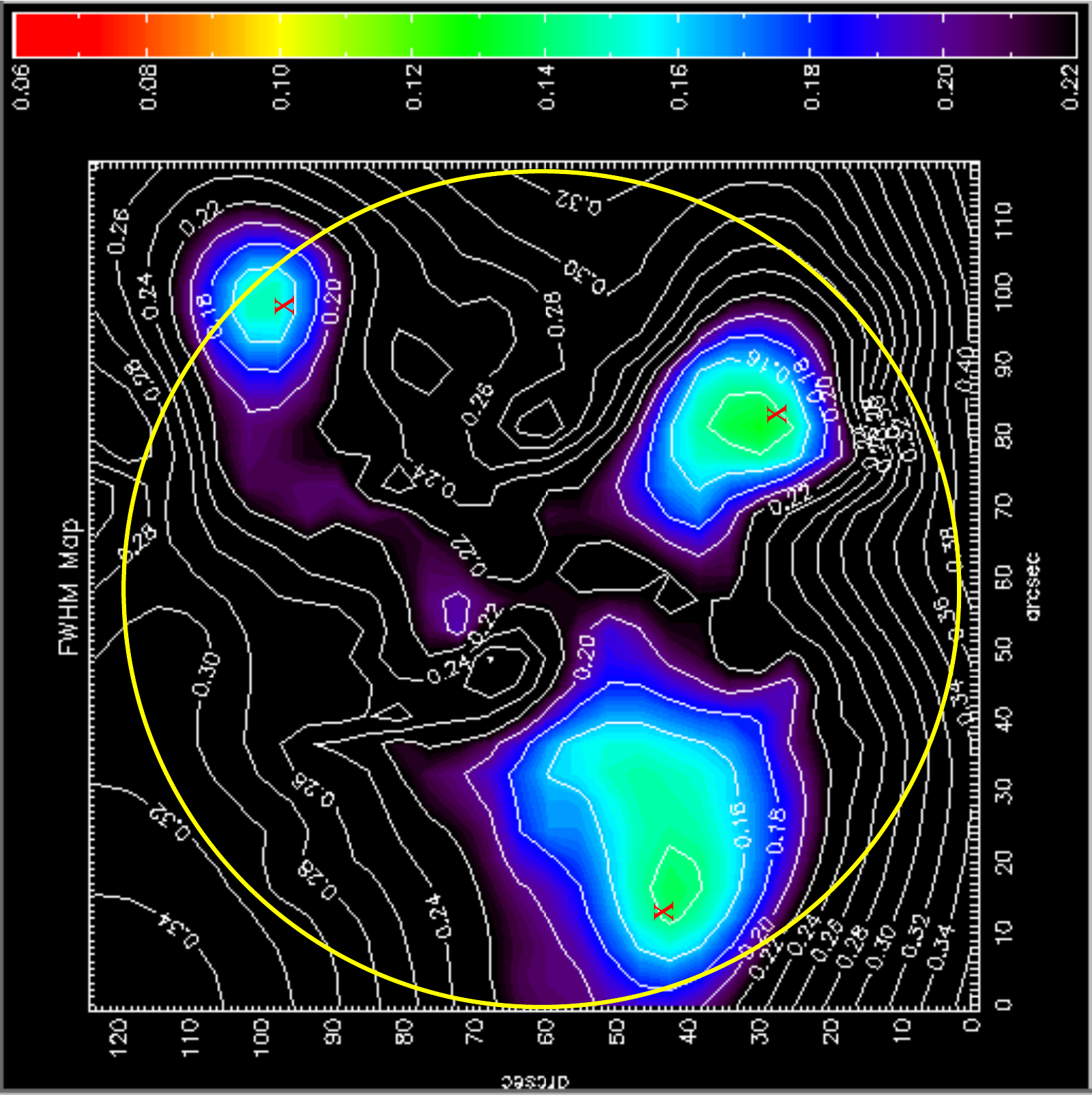}
   \includegraphics[width=\columnwidth,angle=-90]{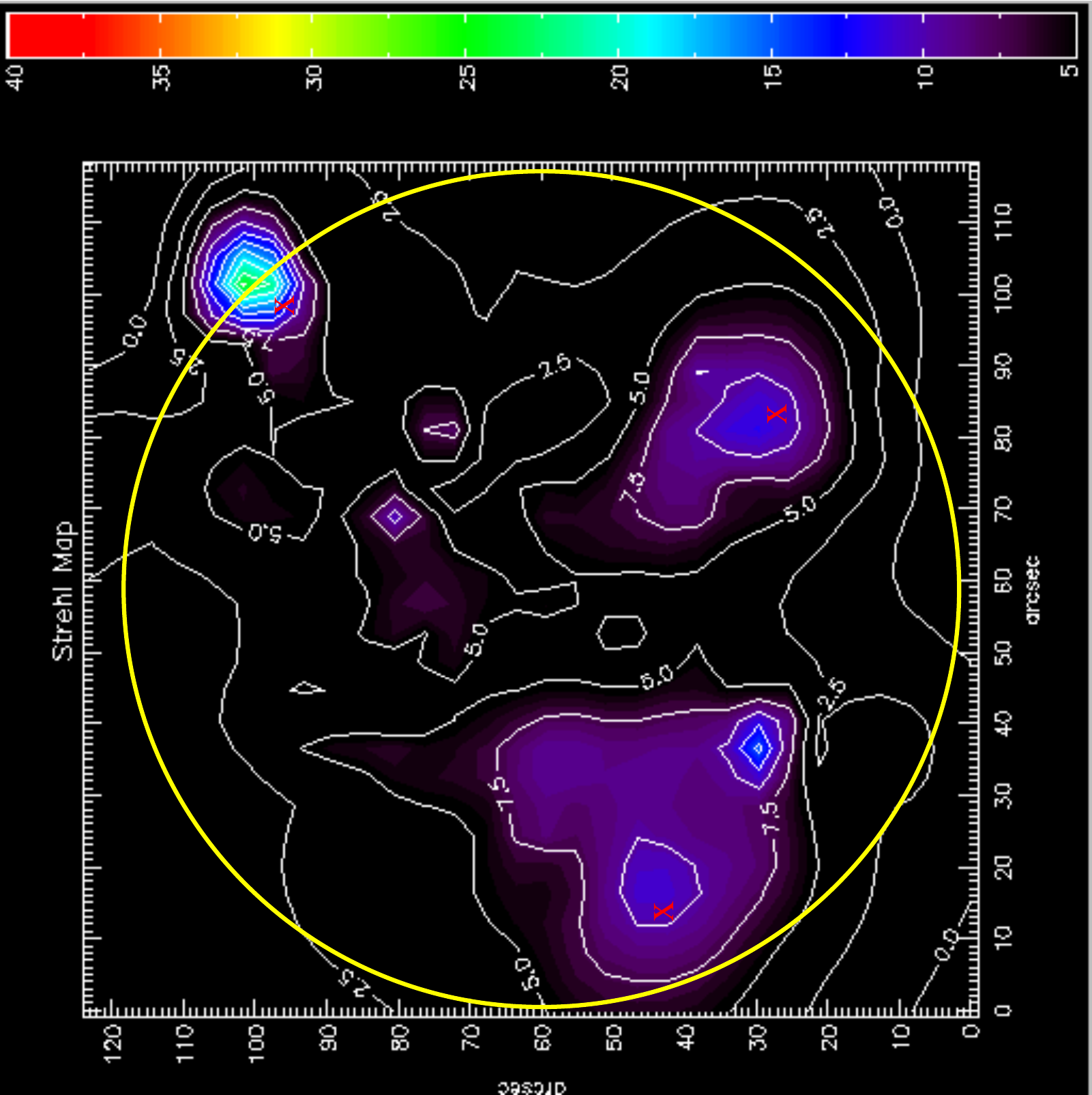}
      \caption{Averaged FWHM and Strehl ratio maps as computed over the MAD field of view
      (yellow circle). The three red crosses show the locations of the NGSs. 
      North is to the top and East to the right.}
   \label{fig: fwhm}
   \end{figure}

Located inside Carina at a distance of 1.5-3.0 kpc,
Tr~14 is an ideal target to search for multiplicity around massive stars
because it contains more than 10 O-type stars and several hundreds of B-type
stars \citep{VBF96}.
Large differences in the distance to Trumpler 14 arise from adopting different 
extinction laws and evolutionary tracks \citep{CRV04}.
Differences in distance can partially account for diverse estimates
of mass and structural parameters.
Its mass was first estimated to be 2000~\msun\ \citep{VBF96}.
 However, the photometry used by these authors barely reached
the turn-on point of the pre-main sequence (PMS), while they extrapolated the mass
assuming a Salpeter initial mass function (IMF).
More recently, \citet{AAV07} used much deeper IR photometry, which
revealed the very rich PMS population and provided a more robust mass estimate
of 9000~\msun.
 \citet{VBF96} reported a core radius of 4.2~pc, while \citet{AAV07}
revised it to 1.14 pc, and detected for the first time a core-halo structure, which is typical of these young clusters \citep[e.g.,][]{BVC04}.  Tr~14 is indeed very young, not yet relaxed and has been forming stars in the last 4 Myr \citep{VBF96}.

   \begin{figure*}
   \centering
   \includegraphics[width=1.5\columnwidth]{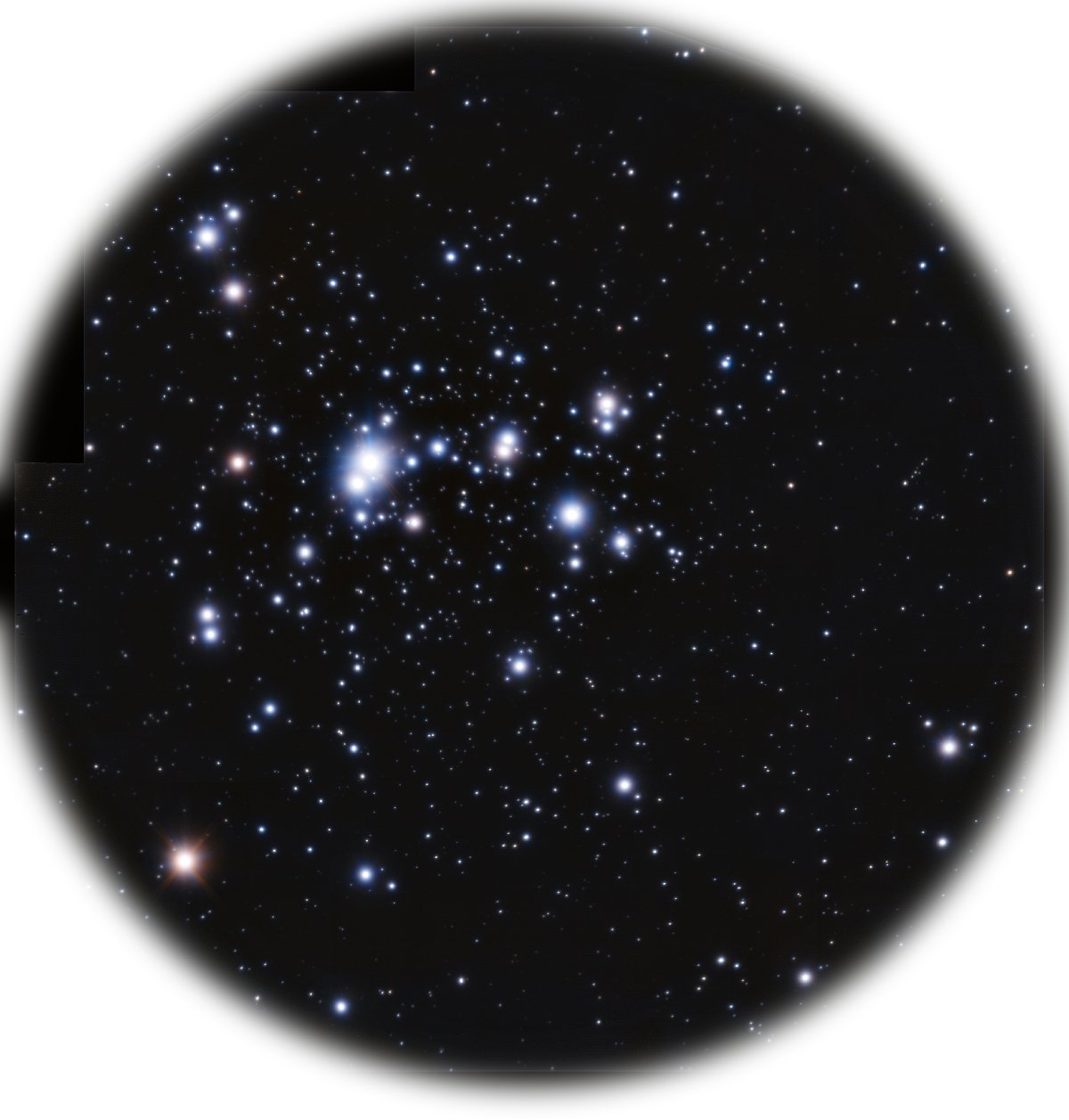}
   \caption{False colour image of the 2\arcmin\ fov of the MAD observations (blue  is $H$ band; red, $K_\mathrm{S}$ band).   North is to the top and East to the left.}
   \label{fig: mosaic}
   \end{figure*}

The layout of the paper is as follows. Sections~\ref{sect: obs} and
\ref{sect: photom}  describe the observations, data reduction and
photometric analysis. Section ~\ref{sect: tr14} presents the NIR
properties of Tr~14 and discusses the cluster structure. 
Section~\ref{sect: comp} analyses the pairing properties in Tr~14.
Section~\ref{sect: bias} describes an artificial star experiment
designed to quantify the detection biases in the vicinity of the bright stars.
It also presents two simple models that generalise the results of the 
artificial star experiment. 
As such, it provides support to the results of this paper.
Finally, Sect.~\ref{sect: mst} investigates the cluster mass segregation status.
and Sect.~\ref{sect: ccl} summarizes our results.

   \begin{figure}
   \centering
   \includegraphics[width=\columnwidth]{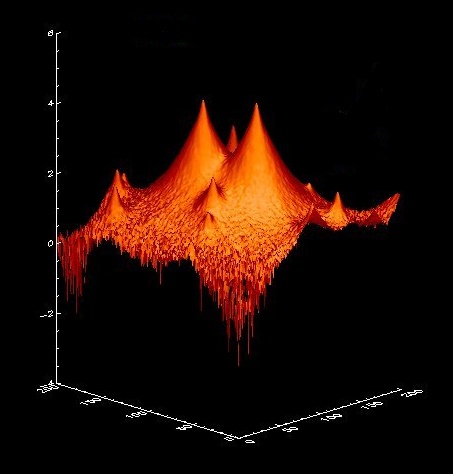}
   \caption{Close-up view of the $K_\mathrm{S}$ band image around selected massive stars. The image size is 200$\times$200 pixels, corresponding to approximately 5.5$\times$5.5\arcsec\ on the sky.}
   \label{fig: psf}
   \end{figure}


\section{Observations and data reduction} \label{sect: obs}
The MAD instrument is an adaptive optics facility aiming at correcting
 for the atmospheric turbulence over a wide field-of-view, and as
 such constitutes a pathfinder experiment for MCAO techniques. Briefly,
MAD relies on three natural guide stars (NGSs)
to improve the image quality (IQ) over a 2\arcmin-diameter fov. 
Optimal correction is reached within the triangle formed by the three
NGSs although some decent correction is still attained outside, mostly
depending on the observing conditions and on the coherence time of
the atmospheric turbulence.

The CAMCAO IR camera images a 57\arcsec$\times$57\arcsec\ region
in the MAD fov and is mounted on a scanning table, so that the full 
2\arcmin-diameter fov can be covered with a  4- or 5-point dither pattern.
The detector used is an Hawaii 2k$\times$2k, yielding an effective
pixel size on sky of 0.028\arcsec. 

Because of the constraints imposed by the geometry and the magnitudes of
the NGSs as well as by the brightness limit of the detector (typically
$K_\mathrm{S}>8$), the Carina clusters turned out to be ideal targets for our
purposes. Combined with the large collecting area of an 8-m class
telescope, MAD was offering a unique opportunity to collect the missing
high-spatial resolution observations to characterize a statistically significant set of massive
stars.

On the night of January 10, 2008 during the second Science
Demonstration (SD) run, the MAD team acquired $H$ and $K_\mathrm{S}$ band observations
of Tr~14. Because of the mentioned
constraints on the choice of the NGSs, the MAD pointing  was offset
from the cluster centre by about 0.4\arcmin\ to the W-SW and a 4-point 
dither pattern was used to cover the (almost) full 2\arcmin\ diameter
fov  (Fig.~\ref{fig: fov}). 
We obtained 28 images for a total exposure time of 28~min on the central field of the cluster (DP\#1) and eight 30sec images in the three remaining DPs. The size of the jitter box was
10\arcsec. We further followed a standard object-sky-object strategy. 
The jitter and dither pattern for the on- and off-target observations
were identical, although the sky observations were obtained without AO correction. The sky field,
 located about 5\arcmin\ SW from the cluster centre, is
one of the few IR-source depleted regions of the neighbourhood.
The journal of the on- and off-target  observations is summarized in Table \ref{tab: diary}.
For each dither position (Col.~1), Cols.~2 and 3 list the coordinates 
of the four-point dither pattern. The detector integration time (DIT) and
the number of repetitions at each jitter position (NDIT) are given in Cols.~4 and 5.
Finally, Cols.~6 and 7 respectively indicate the number of images and the total integration 
time spent on each dither position.

The data were reduced with the {\sc iraf} package. All science and sky frames were dark-subtracted and flatfielded with the calibrations obtained by the SD-team in January 2008. A master-sky was created for each dither position (DP) by taking the median of the corresponding sky images. The few visible sources in the sky images were manually masked out before computing the median sky. This master-sky was subtracted from the individual science images. While the point spread function (PSF) photometry (see Sect.~\ref{sect: photom}) was performed on the individual images, we also combined the images into a 2\arcmin\ diameter mosaic that was used to estimate the overall IQ of the campaign. We note the presence of electronic ghosts offset by $n\times128$~pix from a bright source (with $n=\pm1, \pm2, \pm3, ...$) along the detector reading direction. We thus systematically masked out the corresponding sub-regions. Thanks to the jittering and to the fact that two contiguous CAMCAO quadrants do not have the same read-out orientation, most of the masked zones were recovered when combining the different frames.

Ambient conditions during our observations were as follows.
The $R$ band seeing was
varying between 0.9\arcsec\ and 1.8\arcsec, corresponding to a $K_\mathrm{S}$ band
seeing between 0.7\arcsec\ and 1.5\arcsec. The coherence time was in
the range of 2 to 3~ms. Even though the ambient conditions were clearly
below average for the Paranal site, the MCAO still provided a decent
improvement with an full-width half maximum (FWHM) of the PSF
 of 0.2\arcsec\ over most of the 2\arcmin\ fov.
The corresponding Strehl ratio was estimated in the range of 5-10\%\
 (Fig.~\ref{fig: fwhm}). Figure~\ref{fig: mosaic} displays a false colour montage of our data on Tr~14, while 
Fig.~\ref{fig: psf} presents a close-up view on a 5$\times$5\arcsec\ region, 
with the aim to emphasize the shape and smoothness of the PSF, even on the   
co-added images.
\\


   \begin{figure}
   \centering
   \includegraphics[width=\columnwidth]{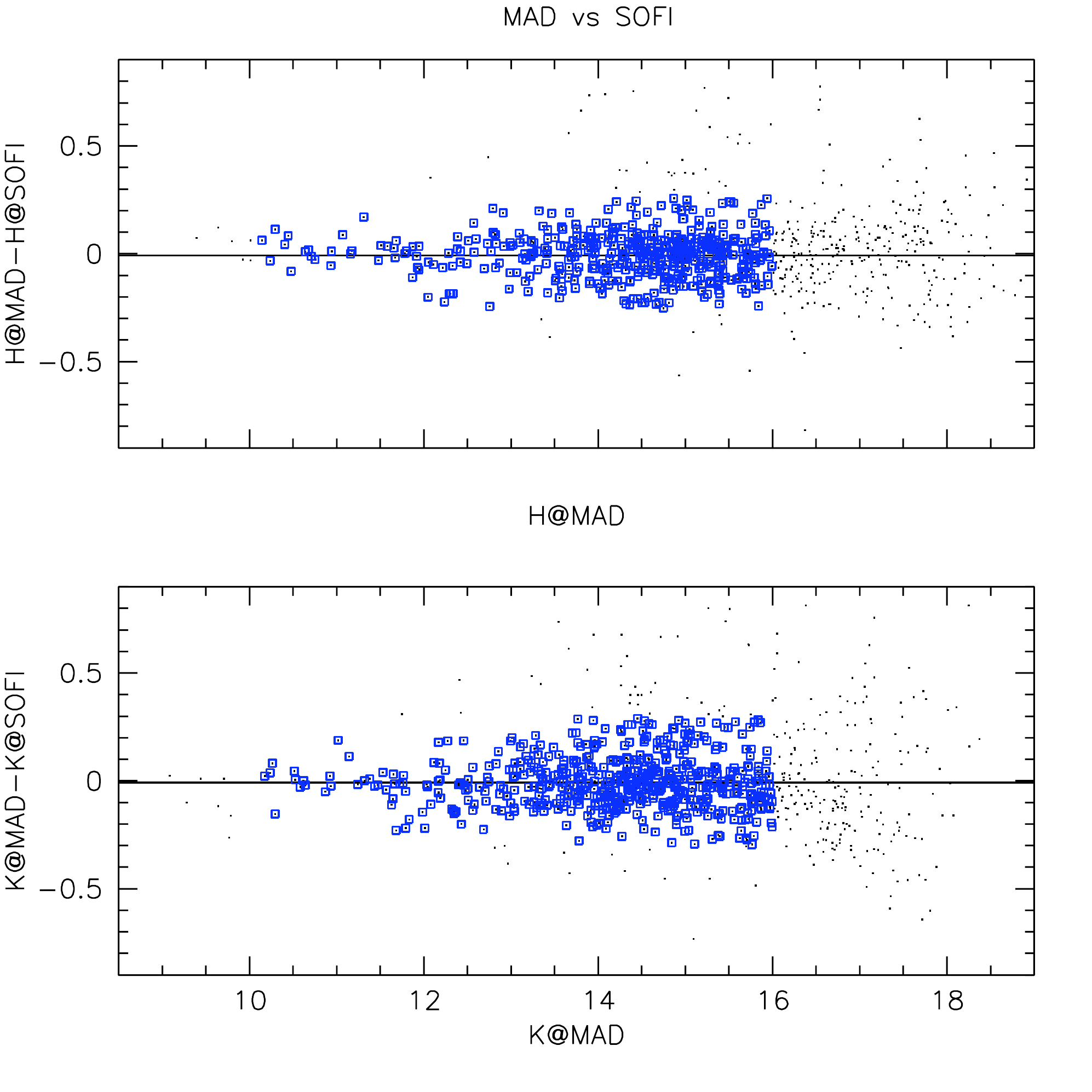}
   \caption{Comparison of the $H$ (upper panel) and $K_\mathrm{S}$  band (lower panel) photometry obtained in this paper (MAD) with that of \citet[SOFI]{AAV07}. The squares indicate the stars used to compute the zero point difference.}
   \label{fig: confsofi}
   \end{figure}

   \begin{figure}
   \centering
   \includegraphics[width=\columnwidth]{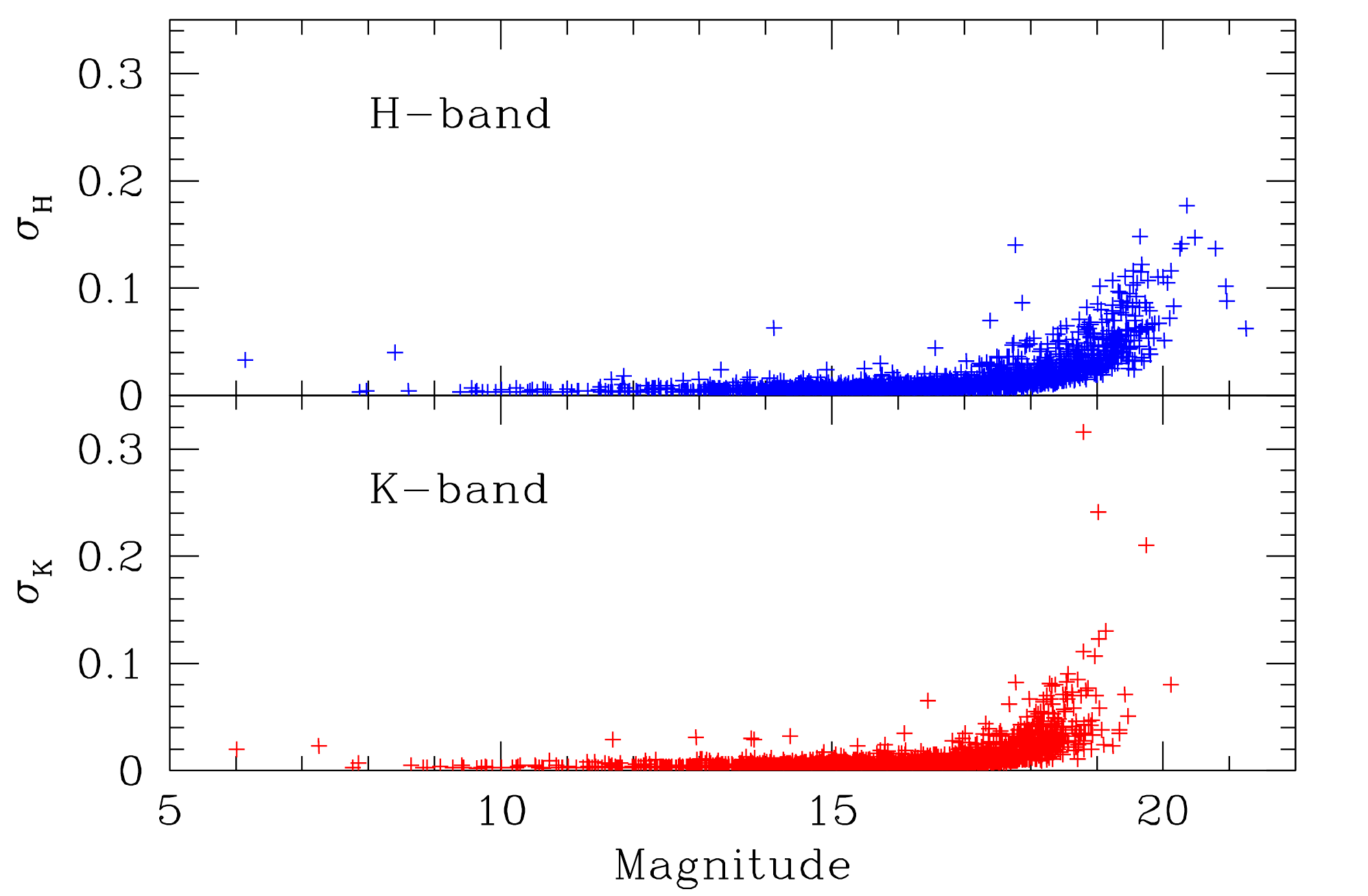}
   \caption{$H$ and $K_\mathrm{S}$ band 1-\s\ error bars on the PSF photometry plotted versus the star magnitude. }
   \label{fig: error}
   \end{figure}

\begin{table*}
\centering
\caption{Photometric catalogue of the sources detected in Tr 14. }
\label{tab: catal}
\begin{tabular}{ccccccc}
\hline
\hline
ID & RA     & DEC.     & $H$    & $\sigma_\mathrm{H}$ &  $K_\mathrm{S}$    & $\sigma_\mathrm{Ks}$  \\
   & HH:MM:SS.sss & $\pm$DD:AM:AS.ss &        &       &        &       \\ 
\hline
 1 & 10:43:46.953 & $-$59:33:18.40 & 12.192 & 0.009 & 11.420 & 0.008 \\ 
 2 & 10:43:47.013 & $-$59:32:43.47 & 16.406 & 0.011 & 15.794 & 0.017 \\ 
 3 & 10:43:47.017 & $-$59:32:51.06 & 15.233 & 0.008 & 14.668 & 0.010 \\ 
 4 & 10:43:47.017 & $-$59:33:09.70 & 11.828 & 0.010 & 10.738 & 0.009 \\ 
 5 & 10:43:47.026 & $-$59:32:42.31 & 11.506 & 0.008 & 11.301 & 0.008 \\ 
 6 & 10:43:47.090 & $-$59:33:24.45 & 17.031 & 0.032 & 16.428 & 0.011 \\ 
 7 & 10:43:47.095 & $-$59:33:17.39 & 15.317 & 0.012 & 14.687 & 0.008 \\ 
 8 & 10:43:47.125 & $-$59:33:12.57 & 17.673 & 0.036 & 16.884 & 0.015 \\ 
 9 & 10:43:47.186 & $-$59:33:21.88 & 14.770 & 0.010 & 13.982 & 0.007 \\ 
10 & 10:43:47.241 & $-$59:33:17.26 & 15.511 & 0.014 & 15.025 & 0.009 \\ 
11 & 10:43:47.309 & $-$59:33:06.09 & 13.557 & 0.012 & 12.963 & 0.010 \\ 
12 & 10:43:47.382 & $-$59:33:41.01 & 14.883 & 0.007 & 14.508 & 0.006 \\ 
13 & 10:43:47.430 & $-$59:33:39.45 & 16.863 & 0.019 & 16.231 & 0.008 \\ 
14 & 10:43:47.464 & $-$59:33:21.92 & 14.537 & 0.012 & 14.070 & 0.008 \\ 
15 & 10:43:47.497 & $-$59:32:40.79 & 14.600 & 0.008 & 13.979 & 0.009 \\ 
16 & 10:43:47.505 & $-$59:33:39.95 & 14.184 & 0.006 & 13.490 & 0.006 \\ 
17 & 10:43:47.516 & $-$59:33:10.58 & 14.125 & 0.012 & 13.773 & 0.009 \\ 
18 & 10:43:47.519 & $-$59:33:26.10 & 17.495 & 0.036 & 16.973 & 0.013 \\ 
19 & 10:43:47.534 & $-$59:33:03.60 & 15.146 & 0.008 & 14.545 & 0.006 \\ 
20 & 10:43:47.539 & $-$59:33:18.65 & 18.158 & 0.043 & 17.211 & 0.019 \\ 
\dots &     \dots &        \dots &  \dots & \dots &  \dots & \dots \\
\hline
\end{tabular}\\
{\sc note:}  The full version of the table is available in the electronic version of the paper or through the CDS: http://cds.u-strasbg.fr/ . 
\end{table*}

\section{PSF photometry} \label{sect: photom}
Stellar photometry was obtained  with the PSF
fitting technique using the well-tested {\sc daophot/allstar/allframe}
\citep{Ste87, Ste94} packages.   The advantage of using {\sc allframe}
is  that  it employs  PSF  photometry on the {\em
individual}  images,  thereby  it  accounts better for the  varying 
near infrared sky and seeing conditions.   The latter have indeed 
a significant impact on the quality of the AO correction and thus of the
actual IQ of the data. The PSF of each individual 
image was generated with a PENNY function that had a quadratic 
dependence on position in the frame, using a selected list of well 
isolated stars.

The calibration of the instrumental $H$ and $K_\mathrm{S}$ MAD data 
was done by direct comparison with the $JHK_\mathrm{S}$ calibrated SOFI 
catalogue of \citet{AAV07}. Of particular importance is the absence 
of  (i) any  colour term  between the  two photometric systems over a 
six-magnitude range (Fig.~\ref{fig: confsofi}), and (ii)  spatial 
systematics between the 4-single 
MAD pointings and  the mean zero-point difference with  respect to the 
SOFI  data. Open squares in Fig.~\ref{fig: confsofi} 
highlight the stars used to estimate the mean 
offset between the two systems, which were selected by applying  a  3$\sigma$ 
clipping around   the  mean  zero-point difference. We obtained  
\begin{eqnarray} 
\Delta~K_\mathrm{S}&=&-0.008\pm0.122 \\ 
\Delta~H&=&-0.007\pm0.101 
\end{eqnarray} 

Taking into account that some objects might be variable and that some 
others were likely not resolved by SOFI, we conclude that there is an almost 
perfect agreement between the two sets of measurements. We therefore continue 
without applying any correction to our photometry. After manually cleaning the 
handful of double entries, our photometric catalogue contains 1955 stars, 
most of them brighter than $K_\mathrm{S}=19$ or $H=20$~mag. 
 A sample of the catalogue is given in  Table~\ref{tab: catal}, where 
Col.~1 indicates our internal identifier, Cols.~2 and 3 give
 the equatorial coordinates and Cols.~4 to 7 provide the $H$ and $K_\mathrm{S}$ band magnitudes and the 1\s\ error bars. 
An overview of the accuracy of the magnitude measurements according to the object brightness is provided in Fig.~\ref{fig: error} . The uncertainties are typically of $\sigma\approx0.005$ for stars brighter than $K_\mathrm{S}=15$~mag ($H=17$~mag resp.) and reach $\sigma\approx0.02$ at $K_\mathrm{S}=18$~mag ($H=19$~mag resp.).

\begin{figure*}
\centering
\includegraphics[width=1.5\columnwidth]{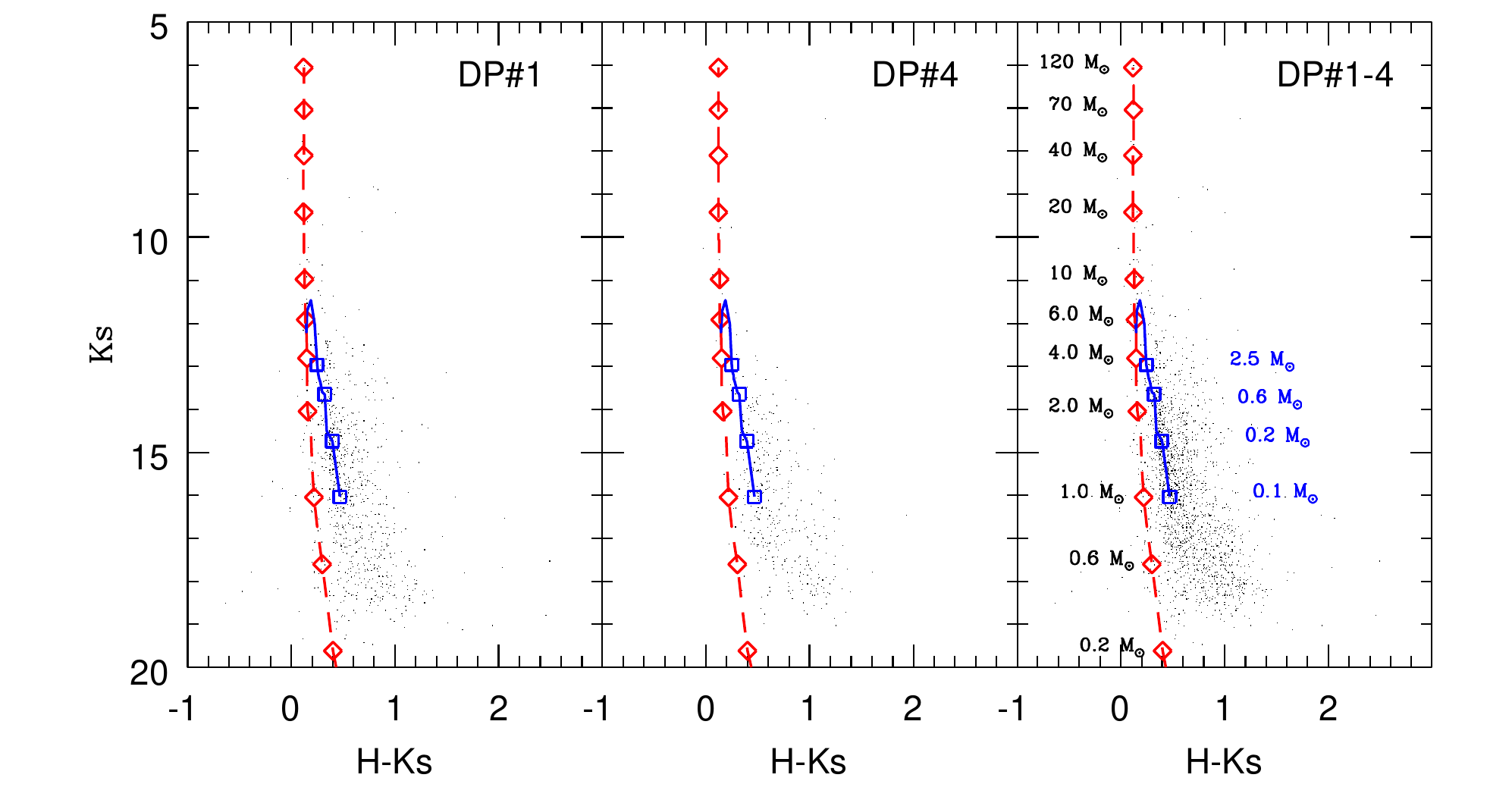}
\caption{{\it Left:} Tr~14 CMD for DP \#1. {\it Middle:} CMD for DP \#4. {\it Right:} complete Tr~14 CMD (DP\#1 to 4). The dashed lines show the MS from \citet{LeS01}  and from \citet{PaS93} for stars with masses $M>6$~\msun\ and $M\le6$~\msun\ respectively. Diamonds and left-hand labels indicate the MS masses. The plain lines show the $\log(age/\mathrm{yr})=5.5$ PMS isochrone from \citeauthor{PaS93} for PMS stars with masses between 0.1 and 6.0~\msun. Squares and right-hand labels indicate the corresponding PMS masses. $DM=12.3$ and $A_\mathrm{V}=2.0$ were adopted as discussed in Sect.~\ref{ssect: cmd}.}
\label{fig: cmd}
\end{figure*}

\section{A MAD view of Tr~14} \label{sect: tr14}

\subsection{Colour-magnitude diagram}\label{ssect: cmd}

Because we only have $H$ and $K_\mathrm{S}$ band observations, our data alone cannot altogether constrain the reddening, the distance and the age of the stellar population. Below, we adopt the results of \citet{CRV04}: $d=2.5\pm0.3$~kpc, $DM=12.3\pm0.2$ and $A_\mathrm{V}=2.0\pm0.13$. 
Figure~\ref{fig: cmd} presents the colour magnitude diagram (CMD) of Tr~14 for dither positions DP \#1 and 4, and for the full fov (DP\#1-4) and compares it with the main sequence (MS) and PMS locations given the adopted cluster distance and reddening. Figure~\ref{fig: cmd} further provides an overview of the light-to-mass conversion scale used in the rest of this paper.

While most of the stars brighter than $K_\mathrm{S}=12$ agree well with the MS of \citet{LeS01}, 
the vast majority of the fainter stars ($K_\mathrm{S}>14$) are still in the PMS stage. A  comparison with the PMS isochrones of \citet{PaS93} suggests a contraction age younger than 1~Myr, and possibly as young as 0.3-0.5~Myr. At this age, the transition between the PMS and the MS occurs for stars with masses in the range of 4~\msun\ to 8~\msun. Although the PMS isochrones are still affected by uncertainties in the colour transformation, our data clearly suggest that the core of Tr~14 has undergone a very recent starburst event, during which most of the low- and intermediate-mass stars in Tr~14 have been formed. Using a larger distance and/or a larger reddening \citep[e.g.,][]{AAV07} would result in even younger ages, which is the reason why we conservatively decided to adopt the former \citeauthor{CRV04} results. 

Without control field observations, we cannot quantitatively estimate the contamination of the CMD by field stars. Yet, the CMD from DP\#4 shows the least structure and provides some qualitative estimate of the maximum contamination suffered by the central part of the cluster. It indicates that the dispersion around the $3\times10^5$~yr isochrone is mostly real. Comparing them with PMS isochrones of different ages,  we estimated the duration of the starburst to be a couple of $10^5$~yr at most.

\begin{figure}
\centering
\includegraphics[width=\columnwidth]{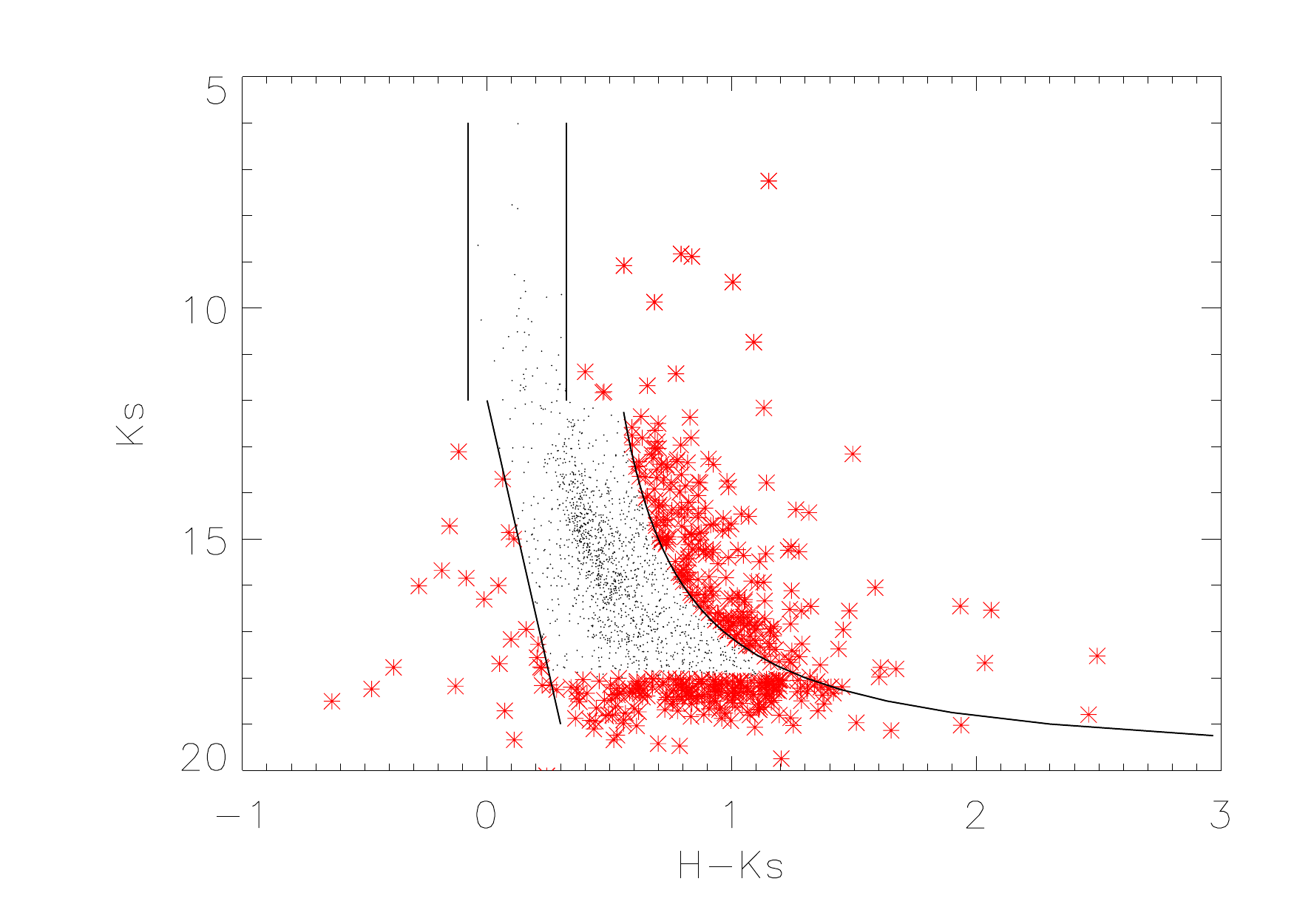}
\caption{Tr~14 CMD. The plain lines delimit the adopted locus of cluster members (see text) while the stars show the objects not considered in Sect.~\ref{sect: comp}.}
\label{fig: sigclip}
\end{figure}


\begin{figure}
\centering
\includegraphics[width=\columnwidth]{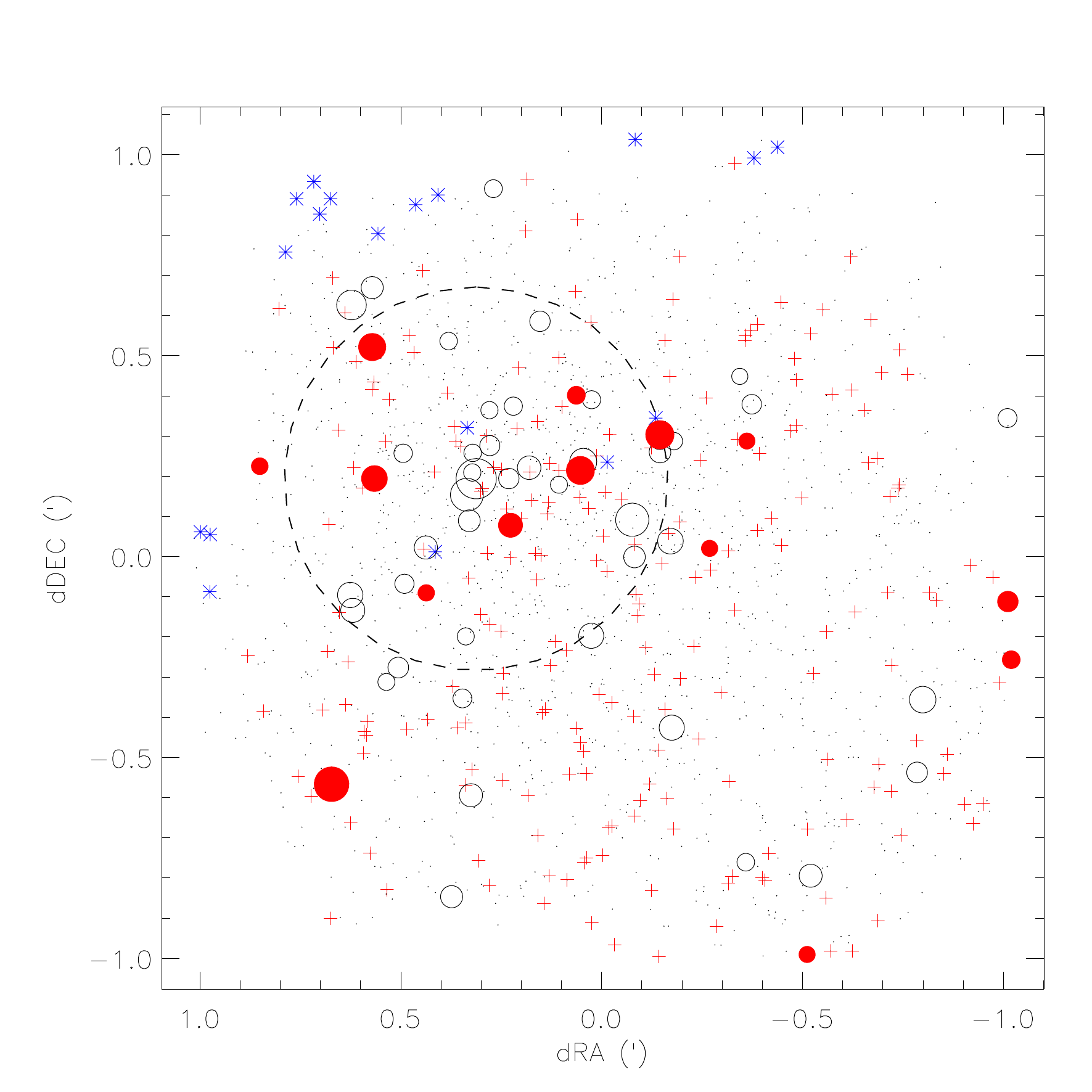}
\caption{Distribution of the reddest and bluest stars in Tr14. The dot and the open circles show the location of the faint ($12<K_\mathrm{S}<18$~mag) and bright ($K_\mathrm{S}>12$~mag) stars respectively. Asterisks and crosses identify the bluest and reddest faint stars, while the filled circles trace the reddest bright stars. For the latter, the symbol size is proportional to the $K_\mathrm{S}$ magnitude. The dashed circle shows a 1\arcmin\ diameter region centered on HD93129A. North is at the top and East to the left.}
\label{fig: bluered}
\end{figure}

\subsection{Foreground/background contamination}\label{sect: clean}
Highly reddened stars seen in the CMD cannot be reproduced by any evolutionary track and are thus deeply embedded cluster members, background objects whose reddened colour results from ISM absorption or foreground cool stars. Similarly, very blue stars in the CMD are unlikely members of Tr~14.

Before statistically estimating the likelihood of a given visual pair to be physically bound (Sect.~\ref{sect: comp}), we first used the current results to clean our catalogue from improbable cluster members. For bright stars ($K_\mathrm{S}>12$~mag), we required the $H-K_\mathrm{S}$ colour to be within 0.2~mag from the expected locus of MS stars with $M>6$~\msol. For stars fainter than $K_\mathrm{S}=12$~mag, we adopted the following relations as the left and right $H-K_\mathrm{S}$ limits in the CMD  (Fig.~\ref{fig: sigclip}):
\begin{eqnarray}
(H-K_\mathrm{S})_\mathrm{left}&=&0.043 (K_\mathrm{S}-12),  \\
(H-K_\mathrm{S})_\mathrm{right}&=&0.3-2.0/(K_\mathrm{S}-20).
\label{eq: red_faint}
\end{eqnarray}
Both relations are obtained as an approximate to the $\pm$2\s\ interval around the typical $H-K_\mathrm{S}$ colour of an 0.3~Myr PMS star for a given $K_\mathrm{S}$ magnitude. Applying those criteria, we excluded 12 bright and 233 faint stars. This represents about 12.5\%\ of our total sample. Below, we also focus on stars brighter than $K_\mathrm{S}=18$~mag. This excludes an additional 198 stars ($\approx$10\%\ of our sample), resulting in a list of 1495 likely members with $K_\mathrm{S}<18$~mag. 

Figure~\ref{fig: bluered} displays the spatial distribution of the rejected blue and red stars. The faint reddest stars agree well with a random spread in the field. The faint bluest ones are mostly located in the North and East edges and suggest larger color uncertainties in that zone. The surface density of the brighter red stars however shows a clear enhancement that correlates with the central part of the cluster. This suggests that the corresponding stars are rather highly reddened objects associated with Tr~14. We acknowledge that we might thus have rejected some members of Tr~14. Preserving the homogeneity of the reddening properties of the bright star sample is however more important and we argue that the resulting few extra rejections will not affect the statistical companionship properties discussed in Sect.~\ref{sect: comp}.

\begin{figure}
\centering
\includegraphics[width=0.9\columnwidth]{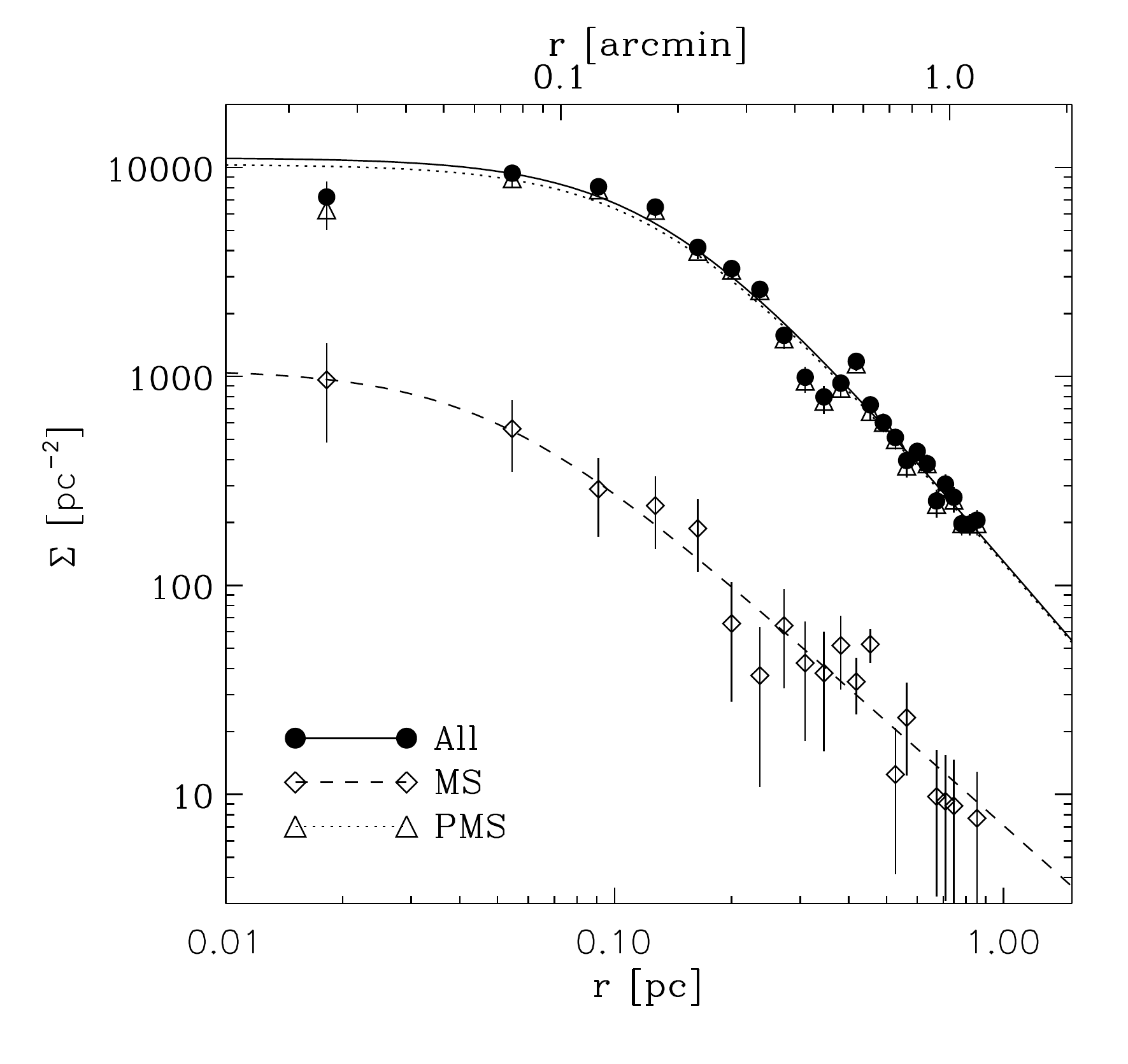}
\includegraphics[width=0.9\columnwidth]{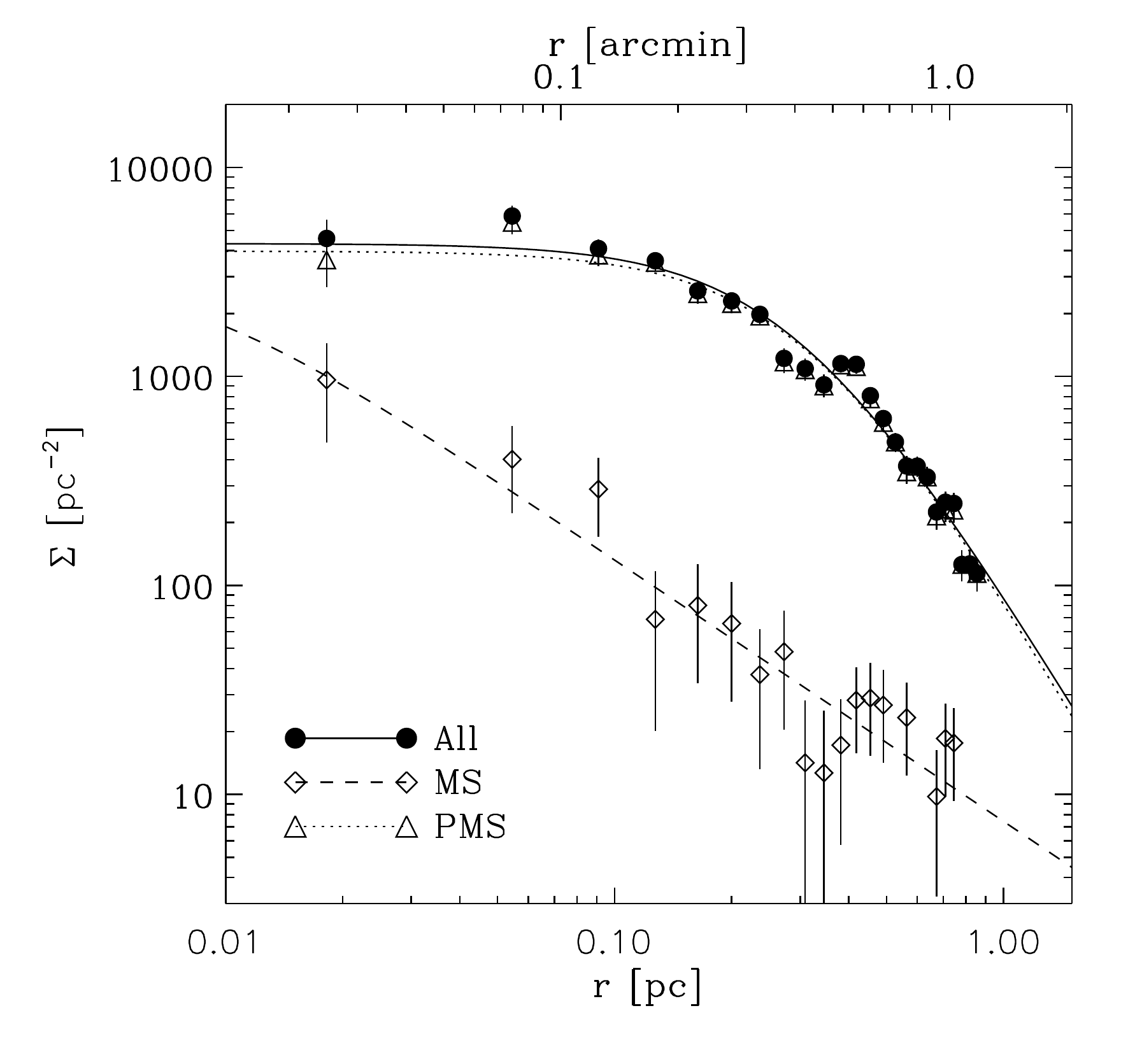}
\caption{Tr~14 surface number density distributions for the full cluster population (circles) and for the PMS (triangles) and MS (diamonds) populations. Upper and lower panels show the density profiles obtained, respectively, before and after applying the colour criteria of Fig.~\ref{fig: sigclip}. In both cases only stars with magnitude $K_\mathrm{S}<18$~mag have been considered. Different lines show the best-fit EFF87 profiles described in Table \ref{tab: eff}.}
\label{fig: eff}
\end{figure}

\begin{table*} 
\centering
\caption{Best-fit EFF87 parameters for different populations in Tr~14. }
\label{tab: eff}
\begin{tabular}{ccccccc}
\hline
\hline
     Population & $K_\mathrm{S}$ range& $\Sigma_0$ [pc$^{-2}$]   &       $a$ [pc]      &    $\gamma$ &  $r_{\rm c}$ [pc] &  $n_0$ [pc$^{-3}$] \\
\hline
\\
\multicolumn{6}{c}{Without colour selection}\\
\\
\vspace*{1mm}  MS &  6--12 &     1077 $\pm$      567 & 0.05 $\pm$ 0.03     & 1.66 $\pm$ 0.27     & $0.06_{-0.03}^{+0.03}$     &   $  9871_{-6376}^{+10497}$  \\
\vspace*{1mm} PMS & 12--18 &    10330 $\pm$      770 & 0.13 $\pm$ 0.01     & 2.17 $\pm$ 0.08     & $0.11_{-0.01}^{+0.01}$     &   $ 40715_{-4520}^{+5019}$  \\
\vspace*{1mm} All &  6--18 &    11116 $\pm$      817 & 0.13 $\pm$ 0.01     & 2.18 $\pm$ 0.08     & $0.11_{-0.01}^{+0.01}$     &   $ 44665_{-4864}^{+5390}$  \\
\\
\multicolumn{6}{c}{With colour selection}\\
\\
\vspace*{1mm}  MS &  6--12 &   3097 $\pm$ 14013 & 0.01 $\pm$ 0.03 & 1.25 $\pm$ 0.39 & $0.01_{-0.05}^{+0.05}$  &    \dots  \\
\vspace*{1mm} PMS & 12--18 &   3977 $\pm$   289 & 0.32 $\pm$ 0.04 & 3.26 $\pm$ 0.19 & $0.23_{-0.03}^{+0.03}$  &    $8348_{-1070}^{+1258}$ \\
\vspace*{1mm} All &  6--18 &   4321 $\pm$   316 & 0.30 $\pm$ 0.03 & 3.10 $\pm$ 0.17 & $0.22_{-0.03}^{+0.03}$  &    $9491_{-1199}^{+1394}$ \\
\hline
\end{tabular}\\
{\sc note:} The uncertainties on $\Sigma_0$, $a$ and $\gamma$ are the 1\s\ error bars.
The uncertainties $r_{\rm c}$ and  $n_0$ were computed using Monte Carlo simulations assuming a Gaussian distribution of the errors on $\Sigma_0$, $a$ and $\gamma$. The quoted values give the 0.68 confidence interval.
\end{table*}

\subsection{Cluster structure}\label{ssect: eff}
Adopting the cluster centre as defined by \citet{AAV07}, we computed the radial profile of the star surface density (Fig.~\ref{fig: eff}).  \citet{AAV07} proposed a core-halo structure with a respective approximate radius of 1\arcmin\ and 5\arcmin. Although our data set covers a limited area, there is no indication of a transition between the two regimes. We can thus consider that our fov is strictly dominated by the core of the cluster. The cluster parameters derived below thus only apply to the core of the core-halo structure.

To better quantify the surface density variations, we fitted \citet[][hereafter EFF87]{EFF87} profiles, better suited for young open clusters than King profiles, to all three populations. Following EFF87, we adopt the notation 
\begin{equation}
\Sigma(r)=\Sigma_0 (1+r^2/a^2)^{-\gamma/2}, \label{eq: eff}
\end{equation} 
where $\Sigma_0$ is the surface number density at the centre.  With the best-fit parameters (Table~\ref{tab: eff}) we also computed the radius, $r_{\rm c}$, where the surface number density drops to half its centre value\footnote{This value is often referred to as the core radius. Because Tr~14 displayed a core-halo structure, $r_\mathrm{c}$ should be understood as the core radius of the core itself.} and the (deprojected) central number density, $n_0$, following Eqs. 22 and 13b of EFF87, respectively. The central density obtained, $n_0=4.5_{-0.5}^{+0.5}\times10^4$~pc$^{-3}$, is very high.

Integrating the surface number density profile to infinity and assuming an average stellar mass of 0.64~\msun\ as suitable for a \citet{Kro01} IMF, our results indicate an asymptotic mass of $\sim4.3^{+3.3}_{-1.5}\times10^3$~\msun, with 30\%\ of the mass within the inner parsec of the profile. This value should be taken as an upper limit to the mass in the cluster core because of the possible contamination of the fov by field stars. Our mass estimate is somewhat lower than the value of $9\times10^3$~\msun\ obtained by \citet{AAV07} based on mass-function considerations. Yet the value of \citeauthor{AAV07} falls within the upper limit of our uncertainties after correcting for the different assumption on the distances. Part of the difference in the mass determination might also be caused by our profile only  probing the core region of the core-halo structure, which likely has a steeper profile than the halo.

 As a first attempt to search for differences between the low- and high-mass star properties, we also computed the density profiles of two sub-populations in Tr~14: the MS stars ($K_\mathrm{S}<12$~mag) and the PMS stars ($12<K_\mathrm{S}<18$~mag).  From Fig.~\ref{fig: eff} (upper panel), the PMS stars occupy a larger core but display a steeper decrease than the MS population. Because the population of the cluster is dominated by low mass stars, there is little difference between the number density profile of the PMS stars and that of the whole cluster. The more massive MS stars ($K_\mathrm{S}<12$~mag) however seem to be slightly more concentrated towards the cluster centre (see also discussion in Sect. ~\ref{sect: mst}). As mentioned earlier, the PMS profile seems to display a steeper slope, but  Table~\ref{tab: eff} reveals that this difference is not very significant (only at the $1.8\sigma$ level). 

As a second step, we also re-computed the density profiles after applying the colour and magnitude criteria defined in the previous section (Sect.~\ref{sect: clean}), thus focussing on the most probable members. The cluster core radius is found to be larger and the profiles display a significantly steeper decrease with radius (Fig.~\ref{fig: eff}, lower panel). The central surface density is also reduced by a factor 2.6 and the deprojected central number density by a factor 4.7 (Table~\ref{tab: eff}).  Following these adjustments, the asymptotic mass of the core is reduced  to $1.4^{+0.4}_{-0.3}\times10^3$~\msun. Because of the sharper slope of the profile, one now finds 75\%\ of the total mass in the inner parsec. Interestingly, the more massive stars show no core-structure, and their density profile is well represented by a simple power-law. This is somewhat comparable to what \citet{CEM10} found for the massive stars in R136.

   \begin{figure}
   \centering
   \includegraphics[width=\columnwidth]{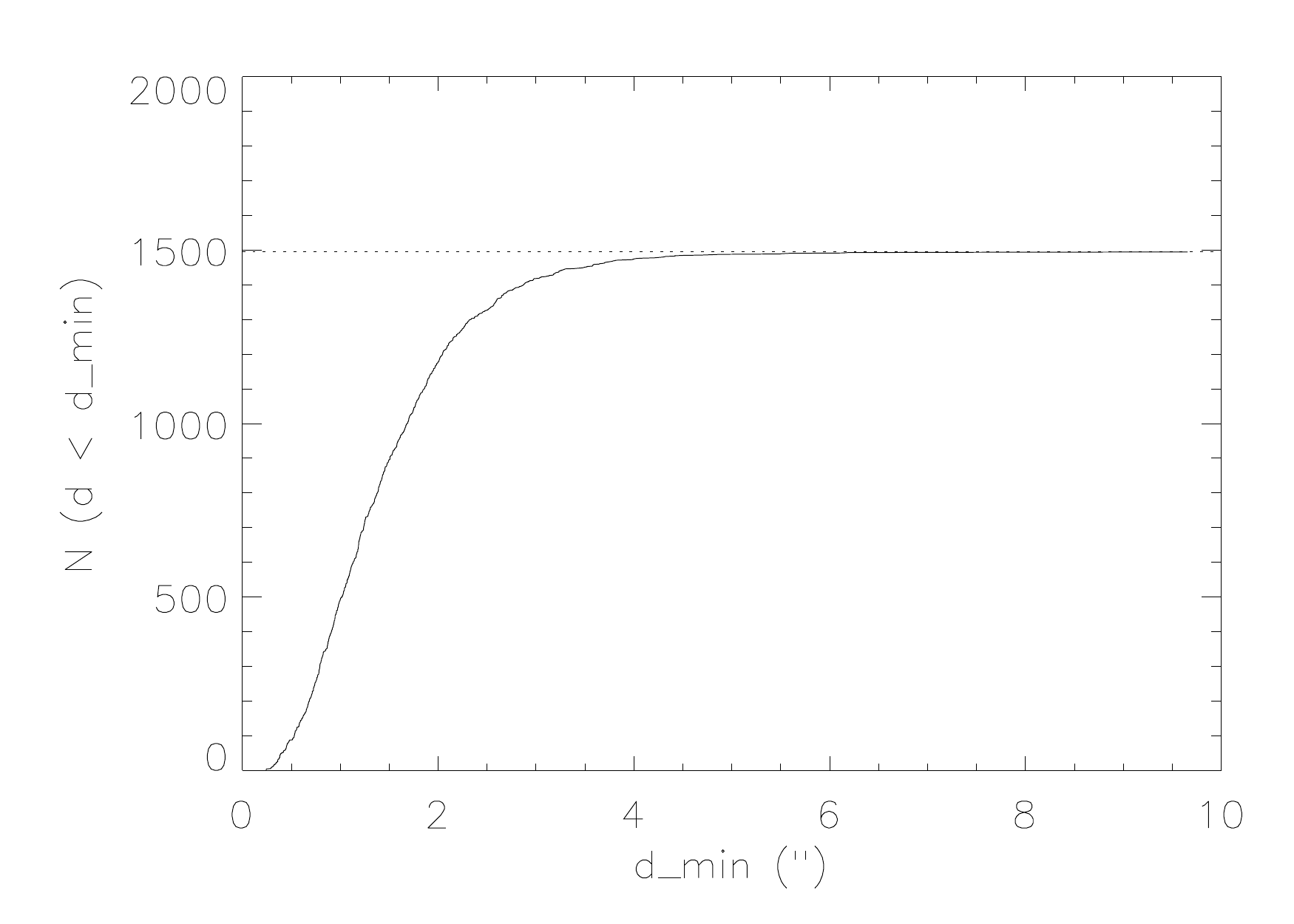}
   \caption{Cumulative distribution of the distance to the closest neighbour ($d_\mathrm{min}$). The horizontal dotted line shows the size of our sample.}
   \label{fig: mindist}
   \end{figure}

   \begin{figure}
   \centering
   \includegraphics[width=\columnwidth]{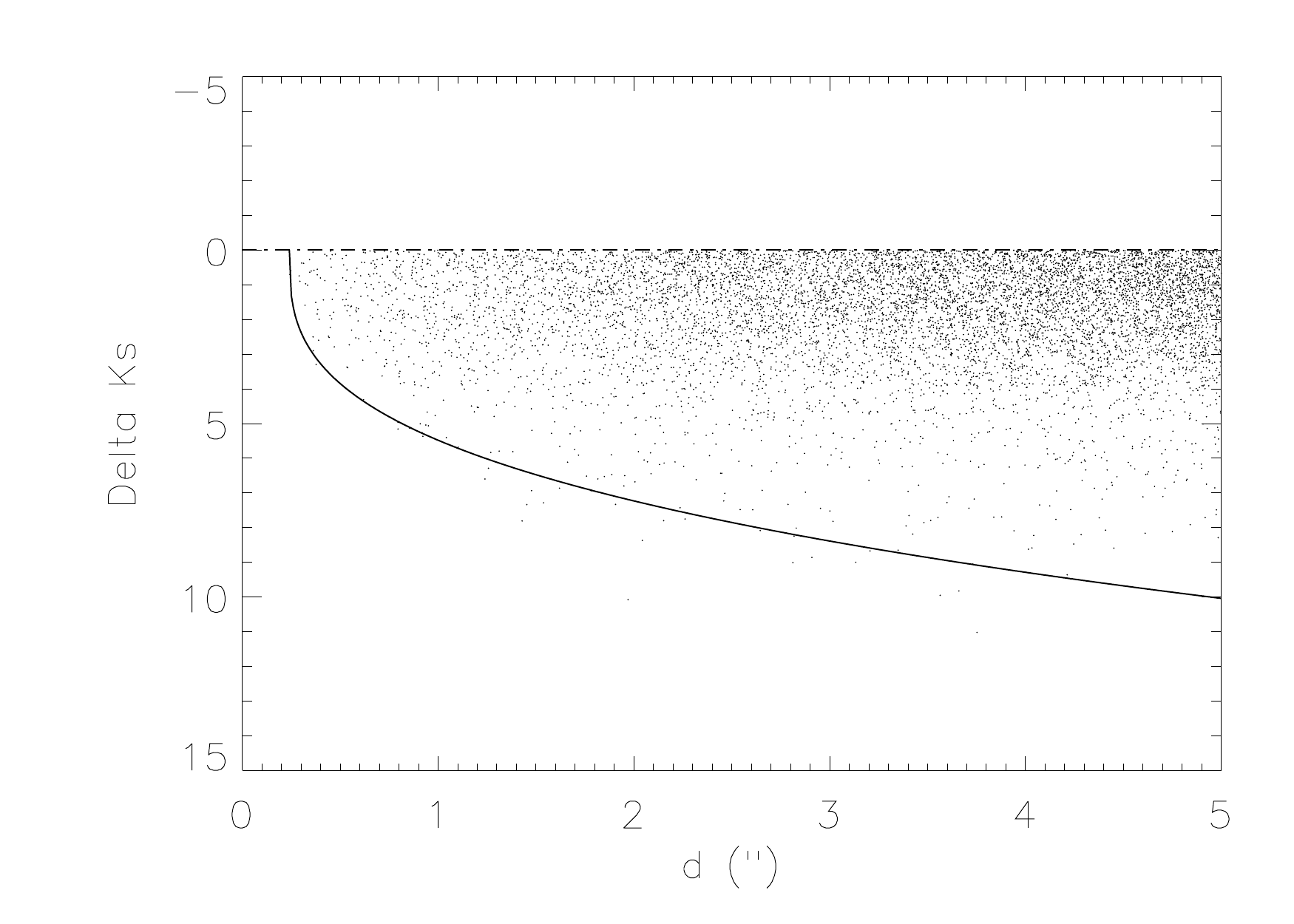}
   \caption{Magnitude difference $\Delta K_\mathrm{S}$ of detected visual pairs as a function of the distance $d$. The plain line is defined by Eq.~\ref{eq: contrast}.}
   \label{fig: dKd}
   \end{figure}

   \begin{figure}
   \centering
   \includegraphics[width=\columnwidth]{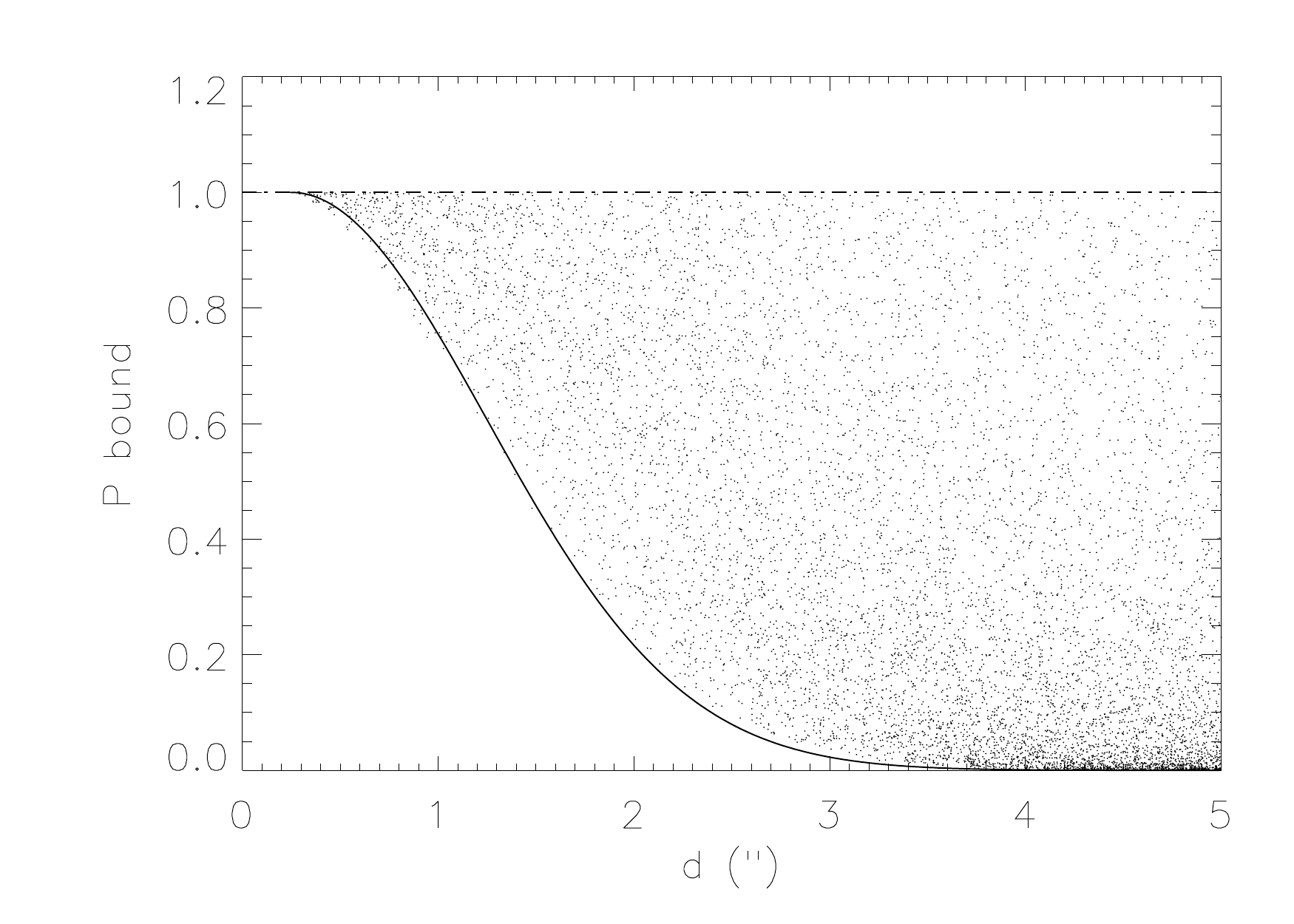}
   \caption{Probability that a pair is physically bound $P_\mathrm{bound}$ plotted vs. the 
 binary separation $d$. The plain curve delimits the locus were no pairs are found and corresponds to $P_\mathrm{bound}=\exp(-(d-0.24)^2/2)$. }
   \label{fig: Pbound}
   \end{figure}

   \begin{figure}
   \centering
   \includegraphics[width=\columnwidth]{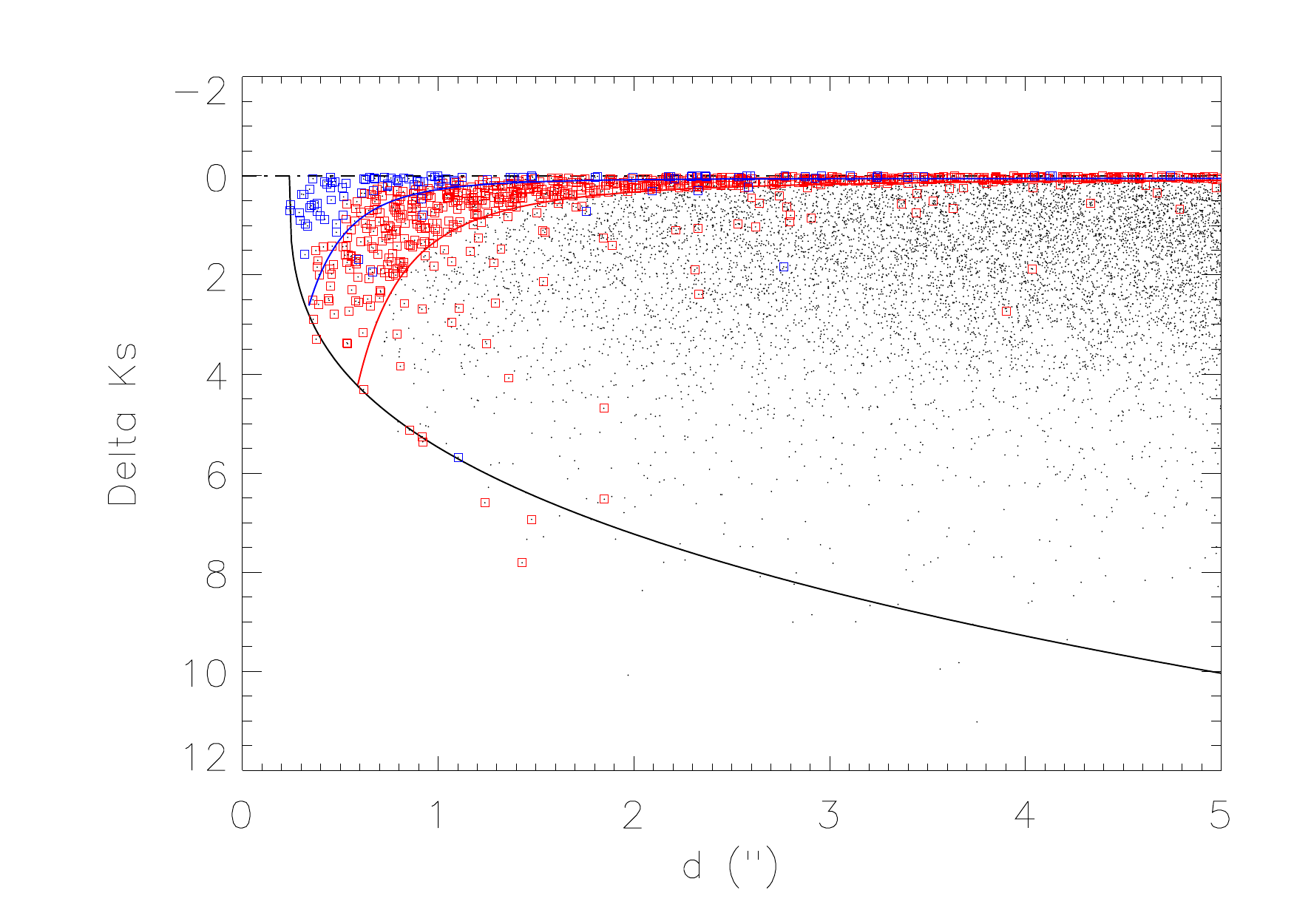}
   \caption{Same as Fig.~\ref{fig: dKd} where the population of likely bound systems has been over-plotted with red ($P_\mathrm{bound}>0.90$) or blue squares ($P_\mathrm{bound}>0.99$). Plain red and blue lines identified the typical  locus of the two populations.}
   \label{fig: dKdp}
   \end{figure}

   \begin{figure}
   \centering
   \includegraphics[width=\columnwidth]{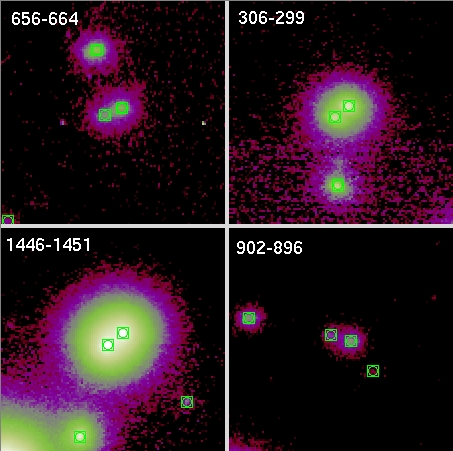}
   \caption{Close-up view in the $K_\mathrm{S}$ band image of the four closest pairs of Table~\ref{tab: pairs}. The displayed regions are 6\arcsec$\times$6\arcsec\ and the separations are all in the range of 0.24\arcsec-0.30\arcsec. Their identifiers as given in Table~\ref{tab: catal} are overlaid on the images.}
   \label{fig: binaries}
   \end{figure}

\section{Companion analysis} \label{sect: comp}

\subsection{General properties} \label{sect: comp_gen}
With about 1500 likely members in a 2\arcmin-diameter fov, the mean surface density is 477 src/arcmin$^2$, or 0.133 src/arcsec$^2$. 
The closest pair detected in our PSF photometry is separated by 0.24\arcsec\ and half the sources have a neighbour at no more than 1.25\arcsec\ (Fig~\ref{fig: mindist}). 
Figure~\ref{fig: dKd} illustrates empirically the maximum reachable flux contrast between two close sources as a function of their separation. The magnitude difference\footnote{Below, the brightest star of a pair is adopted as the primary. We adopt the convention $\Delta K_\mathrm{S}=K_\mathrm{S, sec}-Ks_\mathrm{S, prim}\ge0$, where $K_\mathrm{S, prim}$ and $K_\mathrm{S, sec}$ are the primary and secondary magnitudes respectively.} $\Delta K_\mathrm{S}$ roughly scales as the cubic root of the separation: 
\begin{equation}
\Delta K_\mathrm{S}=6(d-0.24)^{1/3}. \label{eq: contrast}
\end{equation}
Flux contrasts of $\Delta K_\mathrm{S} = 1 / 3 / 5 / 8 $ are reached at separations $d = 0.25 / 0.5 / 1.0 / 2.0$\arcsec\ respectively.

\begin{table}
\centering
\caption{List of likely bound visual pairs ($P_\mathrm{bound}\ge0.99$).}
\label{tab: pairs}
\begin{tabular}{cccccc}
\hline
\hline
ID$_\mathrm{prim}$-ID$_\mathrm{sec}$ & $H_\mathrm{prim}$ & $K_\mathrm{S,prim}$ & $H_\mathrm{sec}$ & $K_\mathrm{S,sec}$ & $d$ (\arcsec)\\ 
\hline
 656- 664 & 17.490 & 16.655 & 18.313 & 17.359 & 0.24 \\ 
 306- 299 & 15.088 & 14.798 & 15.997 & 15.380 & 0.25 \\ 
1446-1451 & 14.097 & 13.323 & 13.803 & 13.560 & 0.28 \\ 
 902- 896 & 17.697 & 17.044 & 18.709 & 17.784 & 0.29 \\ 
1024-1025 & 14.842 & 14.455 & 15.941 & 15.349 & 0.29 \\ 
1091-1092 & 16.165 & 15.342 & 15.979 & 15.459 & 0.29 \\ 
 156- 157 & 17.303 & 16.715 & 17.665 & 17.084 & 0.30 \\ 
1581-1579 & 15.277 & 14.653 & 15.407 & 15.030 & 0.31 \\ 
1488-1491 & 14.205 & 13.941 & 16.044 & 15.530 & 0.32 \\ 
1445-1453 & 16.320 & 16.014 & 17.395 & 16.981 & 0.32 \\ 
1398-1389 & 14.326 & 13.931 & 15.461 & 14.942 & 0.33 \\ 
1152-1149 & 14.867 & 14.498 & 16.254 & 15.229 & 0.34 \\ 
1351-1343 & 12.899 & 12.542 & 15.742 & 14.876 & 0.34 \\ 
 828- 825 & 17.984 & 17.458 & 19.244 & 18.233 & 0.34 \\ 
1780-1779 & 16.527 & 16.052 & 16.829 & 16.322 & 0.34 \\ 
1320-1318 & 15.386 & 15.024 & 16.267 & 15.660 & 0.34 \\ 
 945- 955 & 15.533 & 15.122 & 16.289 & 15.724 & 0.35 \\ 
1163-1155 & 16.471 & 15.241 & 16.277 & 15.674 & 0.35 \\ 
1101-1104 & 12.828 & 12.320 & 15.311 & 14.589 & 0.36 \\ 
1202-1200 & 15.949 & 15.449 & 16.106 & 15.506 & 0.36 \\ 
\dots     &\dots   & \dots  & \dots  & \dots  & \dots\\
\hline
\end{tabular}\\
{\sc note:}  The full version of the table is available in the electronic version of the paper or through the CDS: http://cds.u-strasbg.fr/ . 
\end{table}

\subsection{Chance alignment} \label{sect: comp_chance}

Because of the high source surface density, the number of pairs quickly increases with separation. To quantify the chance that an observed pair results from spurious alignment, we followed the approach of \citet{DSE01}. We define $P_\mathrm{bound}$ as the complementary probability to the one that a given pair $(K_\mathrm{S, prim},K_\mathrm{S, sec}$), with a separation $d$, occurs by chance: 
\begin{equation}
P_\mathrm{bound}=\exp \left(-\sum^{K_\mathrm{S, sec}}_\mathrm{K_\mathrm{S}=K_\mathrm{S, prim}} \frac{W_\mathrm{K_\mathrm{S}}}{\pi R^2} \right),
\end{equation}
 where  $R$ is the field radius, taken as 60\arcsec. $W_\mathrm{K_\mathrm{S}}$ is the actual surface area of a star of magnitude $K_\mathrm{S}$ and is given by 
\begin{equation}
W_\mathrm{K_\mathrm{S}}=\pi \left(d^2-d^2_\mathrm{min} \left(K_\mathrm{S}-K_\mathrm{S, prim}\right)\right),
\end{equation}
 where $d_\mathrm{min}$ is the minimal separation at which a star of magnitude $K_\mathrm{S}$ can be detected in the neighbourhood of a star of magnitude $K_\mathrm{S, prim}$. It is estimated by inverting Eq.~\ref{eq: contrast} thus
\begin{equation}
d_\mathrm{min}(\Delta K_\mathrm{S})= (|\Delta K_\mathrm{S}|/6)^3+0.24. 
\label{eq: dmin}
\end{equation}

As expected, Fig.~\ref{fig: Pbound} shows that the likelihood of finding physically bound pairs, $P_\mathrm{bound}$, decreases steeply with separation. Typically, only pairs with a brightness ratio close to unity or with a separation of less than 0.5\arcsec\ cannot be explained by projection effects (Fig.~\ref{fig: dKdp}). This implies that wider or larger flux contrast systems, if existing, cannot be individually disentangled from the pairs arising by chance alignment along the line of sight. 

Table~\ref{tab: pairs} lists the 150 pairs separated by 5\arcsec\ or less and with $P_\mathrm{bound}\ge0.99$.  The first column indicates the primary and secondary IDs from Table~\ref{tab: catal}. Columns.~2-3 and Cols.~4-5 list the $H$ and $K_\mathrm{S}$ magnitudes of the primary and secondary components, respectively. The separation of the pair is given in Col.~6. The two closest pairs detected have a separation of 0.24\arcsec\  and 0.25\arcsec\ (only 600~AU at the Tr~14 distance) and  $\Delta K_\mathrm{S}\approx$0.6-0.7 mag (Fig.~\ref{fig: binaries}). The closest probable companion to a massive star is a $K_\mathrm{S}=13.2$~mag star at 0.4\arcsec\ from the B1~V star Tr14-19 (pair ID \#1530-1536 in Table~\ref{tab: pairs}). Of the 31 stars brighter than $K_\mathrm{S}=11$~mag, only six have high probability companions in the range of 0.4\arcsec\  to 2.5\arcsec, i.e.\ 19\%. Because of the observational biases discussed and of the filtering criteria applied, the statistical distribution of the selected pairs is not representative of the underlying distribution and will not be discussed further.

   \begin{figure}
   \centering
   \includegraphics[width=\columnwidth]{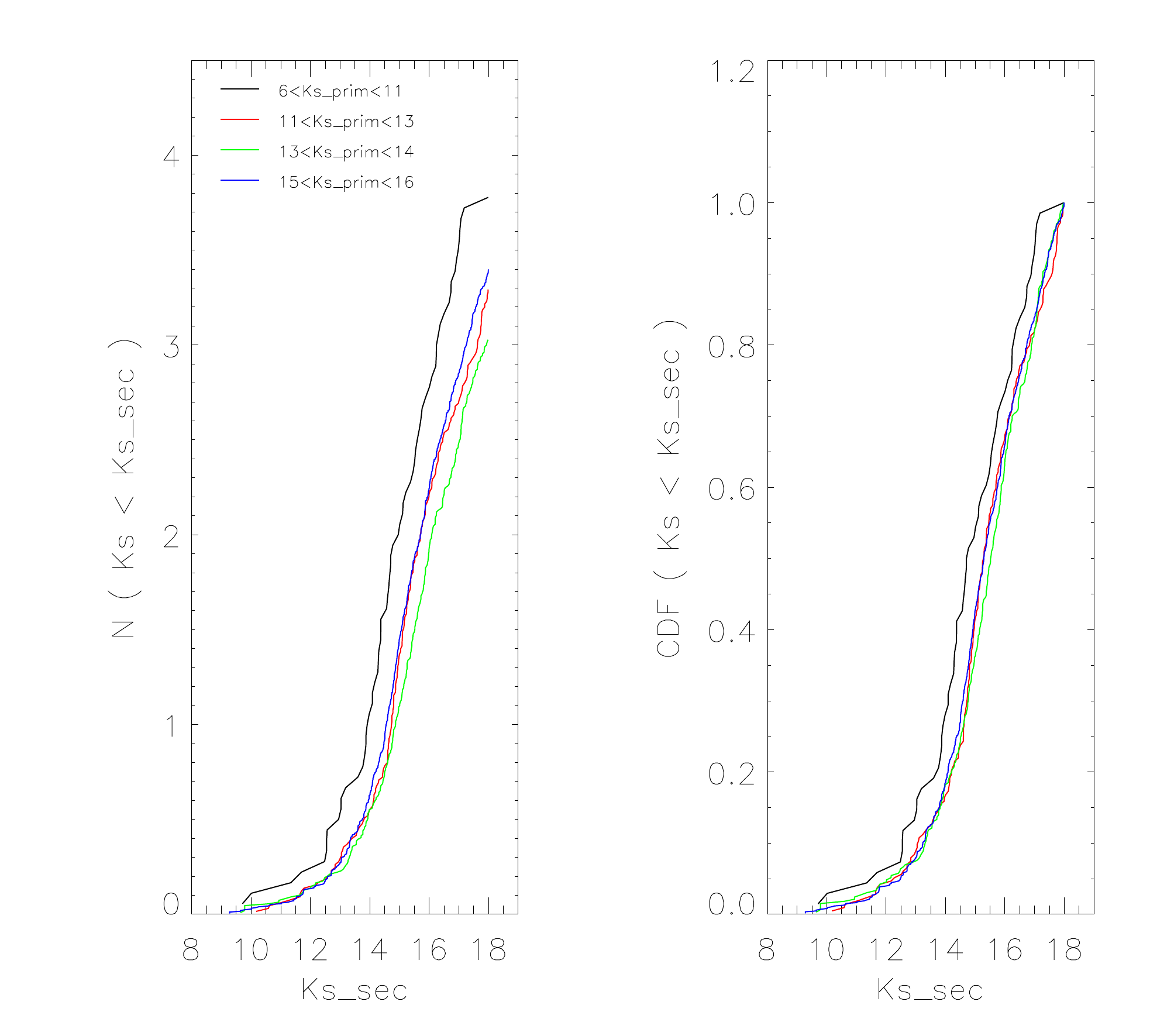}
   \caption{{\it Left:} Average number of companions per star for a given maximum companion magnitude. {\it Right:} Cumulative distribution functions (CDF) of the companion brightness.  The central stars are taken in four ranges as indicated in the upper left-hand legend. Only pairs with separations in the range of 0.5\arcsec-2.5\arcsec\ have been considered.}
   \label{fig: comp}
   \end{figure}


\subsection{Companion frequency} \label{sect: comp_freq}

To search for variations in the companionship properties of different sub-populations, we first computed the average number of companions per star, considering various magnitude ranges both for the central star and for the companions 
(Fig.~\ref{fig: comp}, left panel). 
We applied the same colour and magnitude selections defined in Sect.~\ref{sect: clean}. We further adopted an exclusion radius of 5\arcsec\ between each central star so that each companion is only assigned to one primary, preserving the independence of the different samples.  In this particular case and to allow direct comparison between the different samples, we did not require that $K_\mathrm{S, prim}<K_\mathrm{S, sec}$.  For the central stars, we consider four ranges of magnitudes:
\begin{enumerate}
\item[-] the massive stars : $6<K_\mathrm{S, prim}<11$, corresponding to $M>10$~\msun\ MS stars);
\item[-] the intermediate-mass stars : $11<K_\mathrm{S, prim}<13$, corresponding to $10>M>4$~\msun\ stars;
\item[-] the solar-mass PMS stars : $13<K_\mathrm{S, prim}<14$, corresponding to $2.5>M>0.5$~\msun\ PMS stars;
\item[-] the low-mass PMS stars : $15<K_\mathrm{S, prim}<16$, corresponding to $0.2>M>0.1$~\msun\ PMS stars;
\end{enumerate}
where the isochrones of Fig.~\ref{fig: cmd} have been adopted as guidelines for the mass conversion.
The companion magnitude was chosen in the range of $9.5<K_\mathrm{S, sec}<18$.  
Finally, we restrained our comparison to the 0.5\arcsec-2.5\arcsec\ separation regime. 
As we will show below (Sect.~\ref{sect: artif}), the observational biases have a very limited impact on our results in that range.  

Under those assumptions, we found that massive MS stars have on average $3.8\pm0.5$ companions, while solar-mass PMS stars have $3.0\pm0.2$ companions. The number of companions of intermediate-mass stars and of low-mass PMS stars are not significantly different from one another. Most of the difference is however found for $K_\mathrm{S, sec}<17$: $3.5\pm0.4$ companions for MS stars against $2.5\pm0.2$ for lower mass stars. The difference is thus significant at the 2.5$\sigma$. This corresponds to a rejection of the null hypothesis that massive and lower-mass stars have the same number of companions with a significance level better than 0.01.  Because this is seen against the observational biases (see Sect.~\ref{sect: bias_comp}), this result is likely to be even more significant.

   \begin{figure}
   \centering
   \includegraphics[width=\columnwidth]{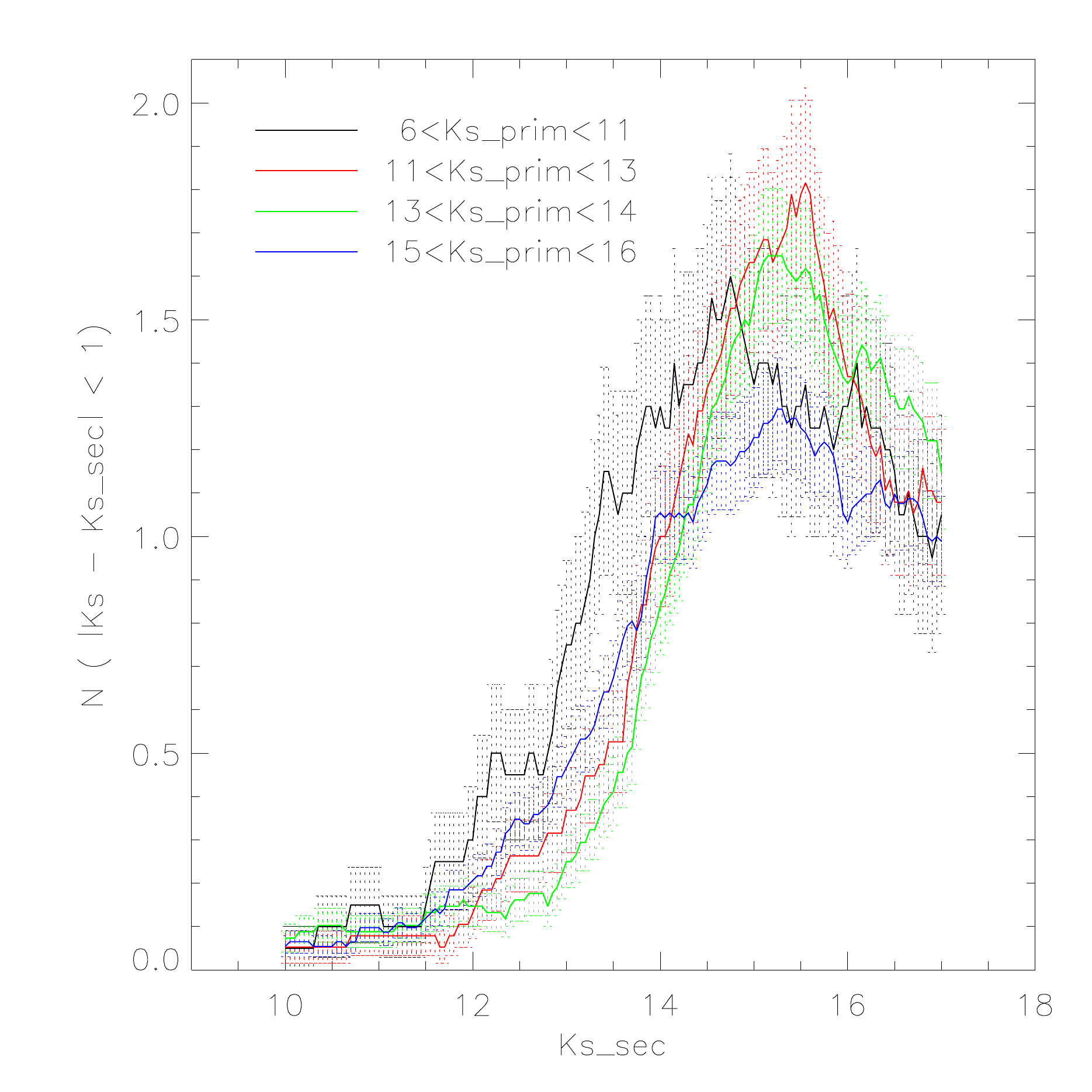}
   \caption{Average number of companions per star as a function of the companion brightness.  The central stars are taken in four ranges as indicated in the upper left-hand legend. A moving average with a 2~mag bin has been used and the considered separation range is 0.5\arcsec-2.5\arcsec. The various envelopes give the 1$\sigma$ error bars.}
   \label{fig: mag_dist}
   \end{figure}

\subsection{Magnitude distribution} \label{sect: comp_mag}
We computed the distributions of the companion magnitudes for the various stellar populations considered above using a 2-mag wide moving average along the companion magnitude axis (Fig.~\ref{fig:  mag_dist}). 
Most of the differences between the distribution functions of the massive stars and of the lower-mass stars result from the range $K_\mathrm{S, sec}< 14$. The more massive stars display thus about twice as  many solar-mass companions as the lower mass stars. The situation is reverse for fainter, low-mass PMS companions, where the lower-mass stars tend to have more companions. The latter effect can however result from the difficulty to detect extremely faint stars in the wings of the brightest stars. 
To allow for quantitative statistical testing, we also built the cumulative distribution functions (CDF) of the companion brightness (Fig.~\ref{fig: comp}, right panel).  Using a two-sided Kolmogorov-Smirnov (KS) test, one can reject at the 2\s\ level the null hypothesis that the high-mass and the solar-mass stars share the same companion CDF.

   \begin{figure}
   \centering
   \includegraphics[width=\columnwidth]{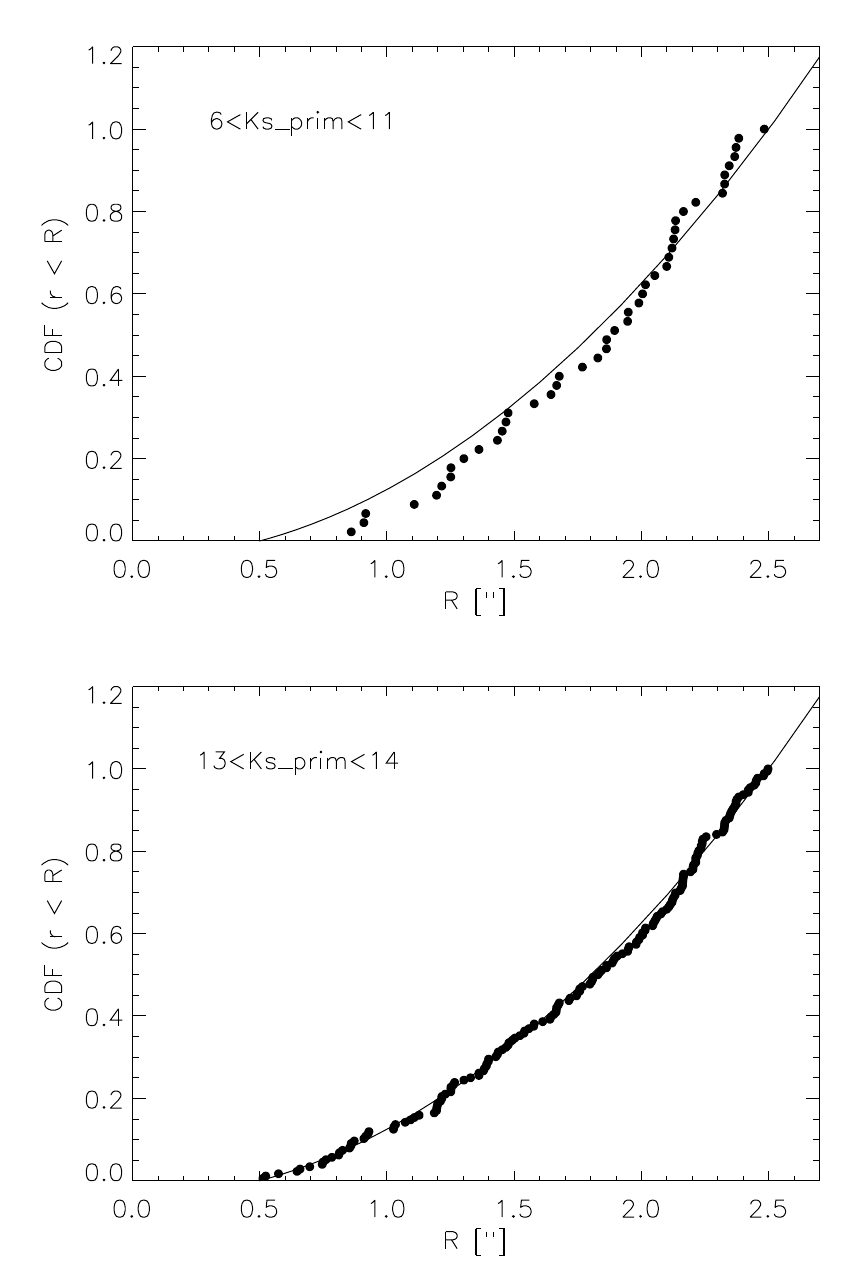}
   \caption{{\it Upper panel:} growth curve for the companions of  massive stars. The plain line indicates the expected distribution for a uniform distribution of the companions in the field of view. {\it Lower panel:}  same as upper panel for the companions of solar-mass stars.}
   \label{fig: growth}
   \end{figure}

\subsection{Companion spatial distribution} \label{sect: comp_spatial}
To investigate the spatial distribution of the companions, we  built the growth curves of the number of companions as a function of the separation. For the curves to be more robust against low number statistics, we concentrated on the total growth curve of a given stellar population rather than on the growth curve of individual targets. We further limited the companions to masses above 0.1~\msol, which roughly corresponds to a magnitude limit of $K_\mathrm{S}=16$~mag and, as in the previous paragraph, we restrained our comparison to the 0.5-2.5\arcsec\ separation regime. 

Figure~\ref{fig: growth} compares the companion growth curves around high-mass and around solar-mass stars with the theoretical distribution expected from random association with an underlying uniform distribution across the field. On the one hand,  massive stars seem to have their companions statistically further away than expected from a uniform repartition. On the other hand, the growth curve of solar-mass PMS stars follows the expected trend  from random association.  Yet in both cases a KS test does not allow us to reject the null hypothesis that both realisations are compatible with the uniform distribution in the considered separation range. Similarly, a two-sided KS test does not allow us to claim that both growth curves are different from one  another.

\subsection{Summary} \label{sect: comp_summary}
To summarize the results of this section, the  closest pair of detected stars in our data is separated by 0.24\arcsec\ ($\sim$600~AU), in good agreement with the IQ of Sect.~\ref{sect: obs}. Equation~\ref{eq: contrast} gives an empirical estimate of the maximum contrast achieved as a function of the separation {Sect.~\ref{sect: comp_gen}. The pairing properties are well described by chance alignment except for the closest pairs ($d<0.5$~\arcsec) and for the pairs with similar magnitude components (Fig.~\ref{fig: dKdp}), yet we identified 150 likely bound pairs (Sect.~\ref{sect: comp_chance}). Massive stars further tend to have more companions than lower-mass stars (Sect.~\ref{sect: comp_freq}). Those companions are brighter on average, thus more massive (Sect.~\ref{sect: comp_mag}). The spatial distribution of the companions of massive stars is however not significantly different from those of PMS star companions (Sect.~\ref{sect: comp_spatial}).  The significance of those results is better than 2\s\ but no better than 3\s, and remains thus limited.   The situation is however reminiscent of the case of NGC~6611 where \citet{DSE01} found that massive stars are more likely to have bound companions compared to solar-mass stars. For Tr~14, the fov is definitely more crowded and probably more heavily contaminated, which seriously complicates both the companionship analysis and the interpretation of the results.


   \begin{figure}
   \centering
   \includegraphics[width=\columnwidth]{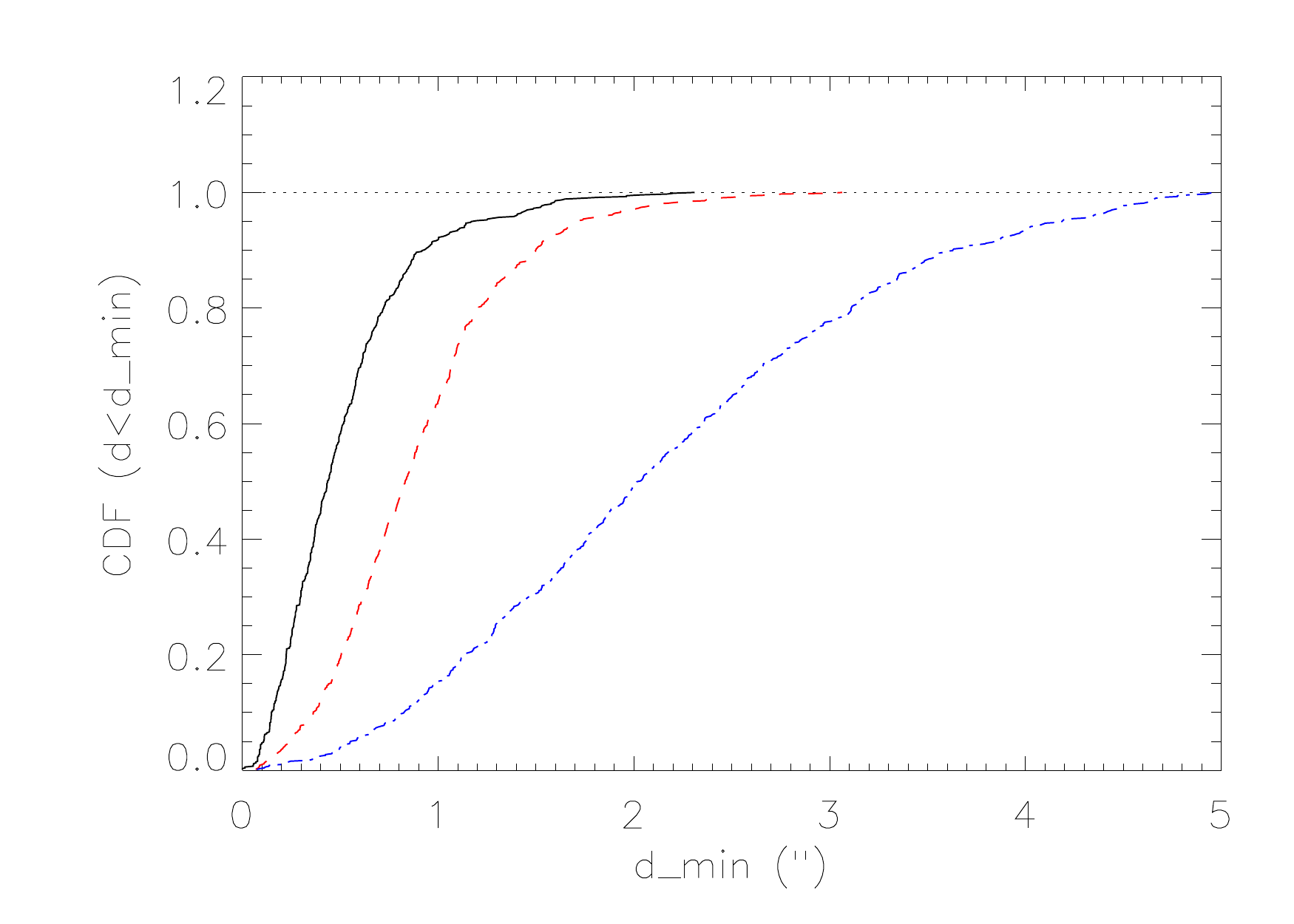}
   \caption{Cumulative distribution functions (CDFs) of the distance from each artificial 
star to (i) the closest artificial star (plain curve), (ii) the closest source in field 
(dashed line), (iii)  the closest bright star with $K_\mathrm{S}<11$ (dashed-dotted line).}
   \label{fig: artif_dmin}
   \end{figure}

   \begin{figure}
   \centering
   \includegraphics[width=\columnwidth]{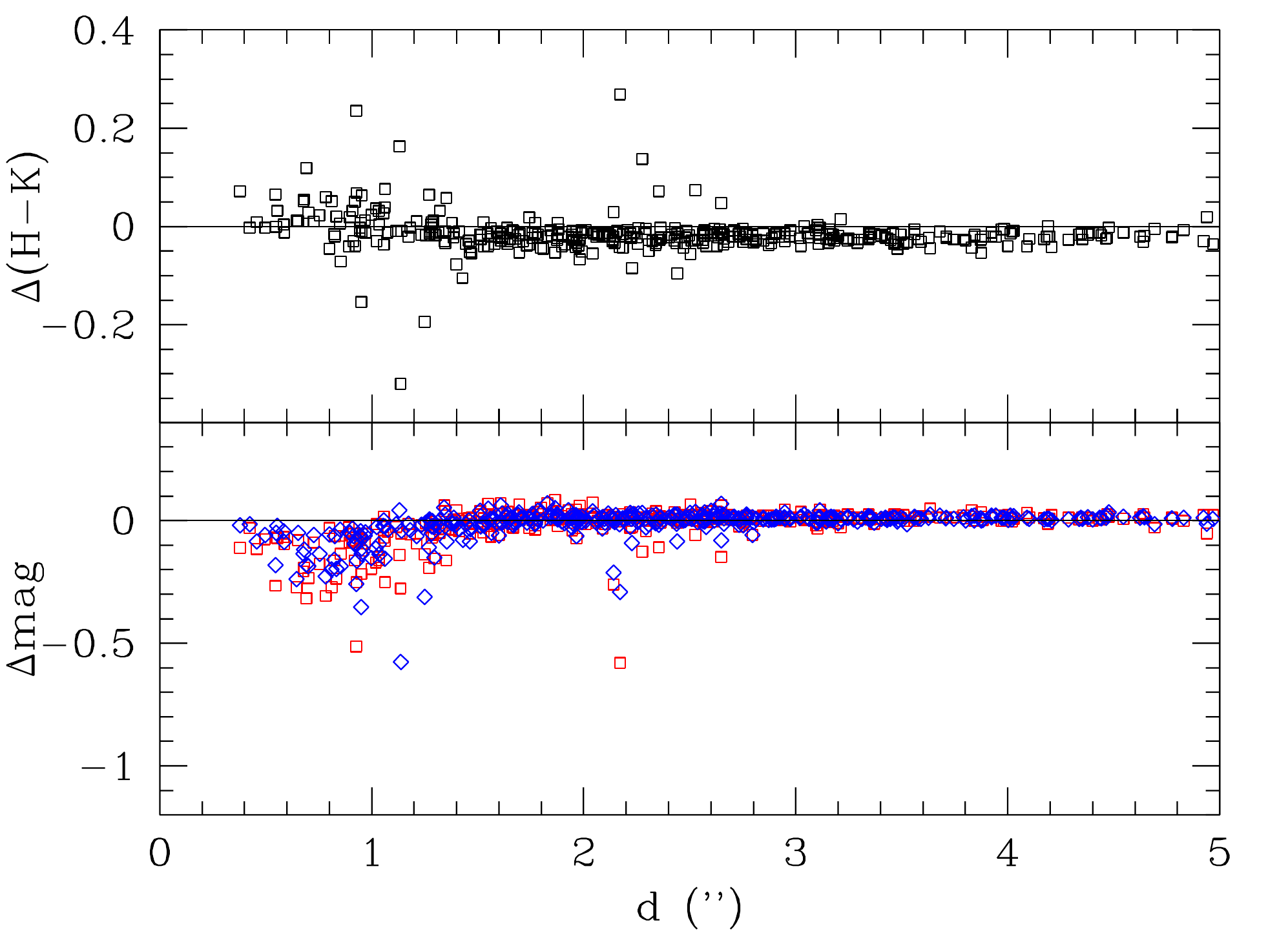}
   \caption{{\it Lower panel:} Comparison between the input and output magnitudes for the artificial star experiment as a function of the separation the to closest bright star in the fov ($K_\mathrm{S}<11$). The squares and diamonds respectively give $\Delta mag=K_\mathrm{S, out}-K_\mathrm{S, in}$ and $H_\mathrm{out}-H_\mathrm{in}$. {\it Upper panel:} same as lower panel for the colour of the artificial stars : $\Delta (H-K)=(H-K_\mathrm{S})_\mathrm{out}-(H-K_\mathrm{S})_\mathrm{in}$.}
   \label{fig: artif}
   \end{figure}

   \begin{figure}
   \centering
   \includegraphics[width=\columnwidth]{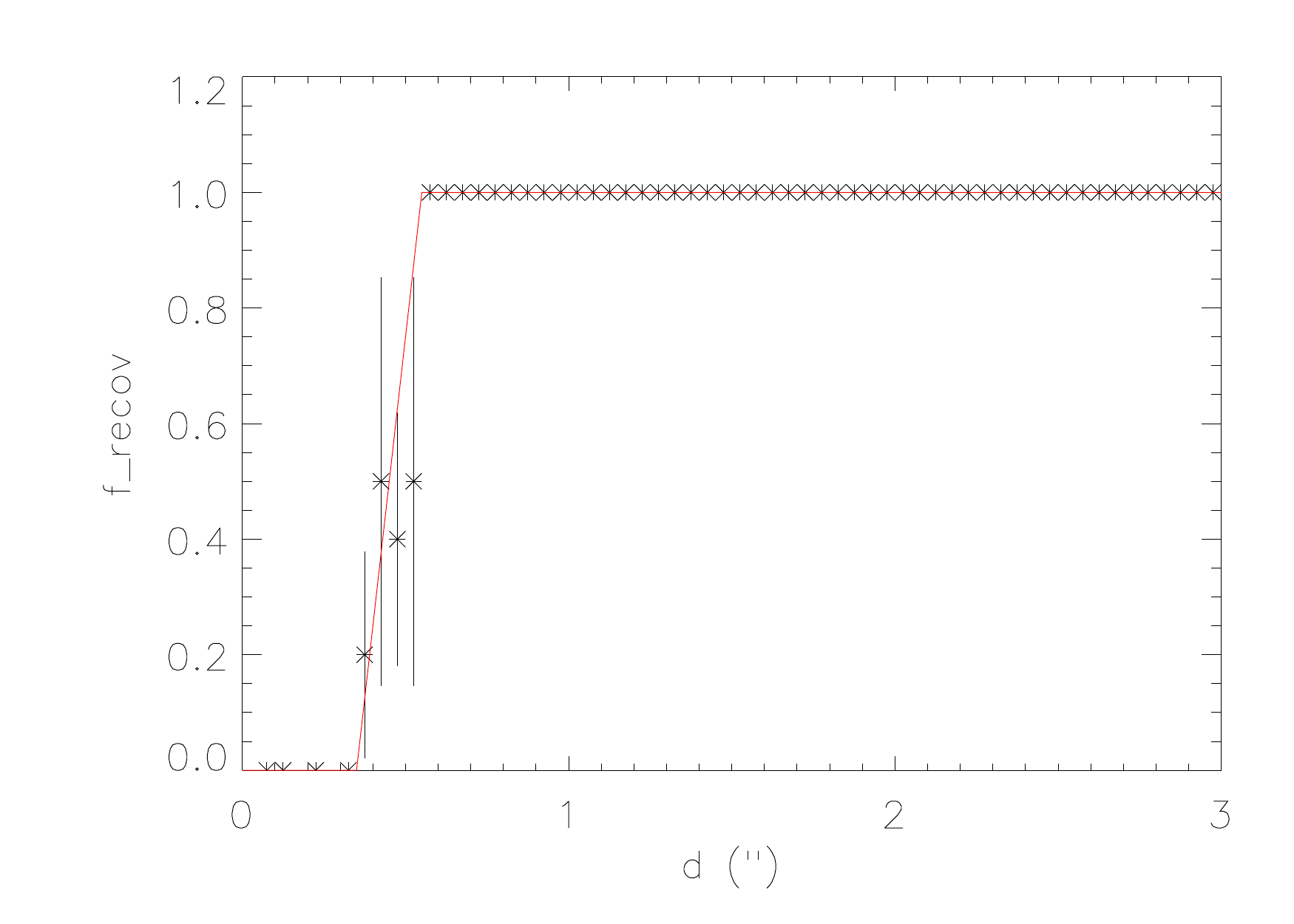}
   \caption{Artificial star recovery fraction as a function of the distance $d$ to a bright neighbour ($K_\mathrm{S}<11$). The plain line shows a  linear interpolation between the detection and non-detection regime and is given by $f_\mathrm{recov}=(d-0.35)/0.2$, for $d$ in the range 0.35--0.55\arcsec.}
   \label{fig: recov}
   \end{figure}

   \begin{figure}
   \centering
   \includegraphics[width=\columnwidth]{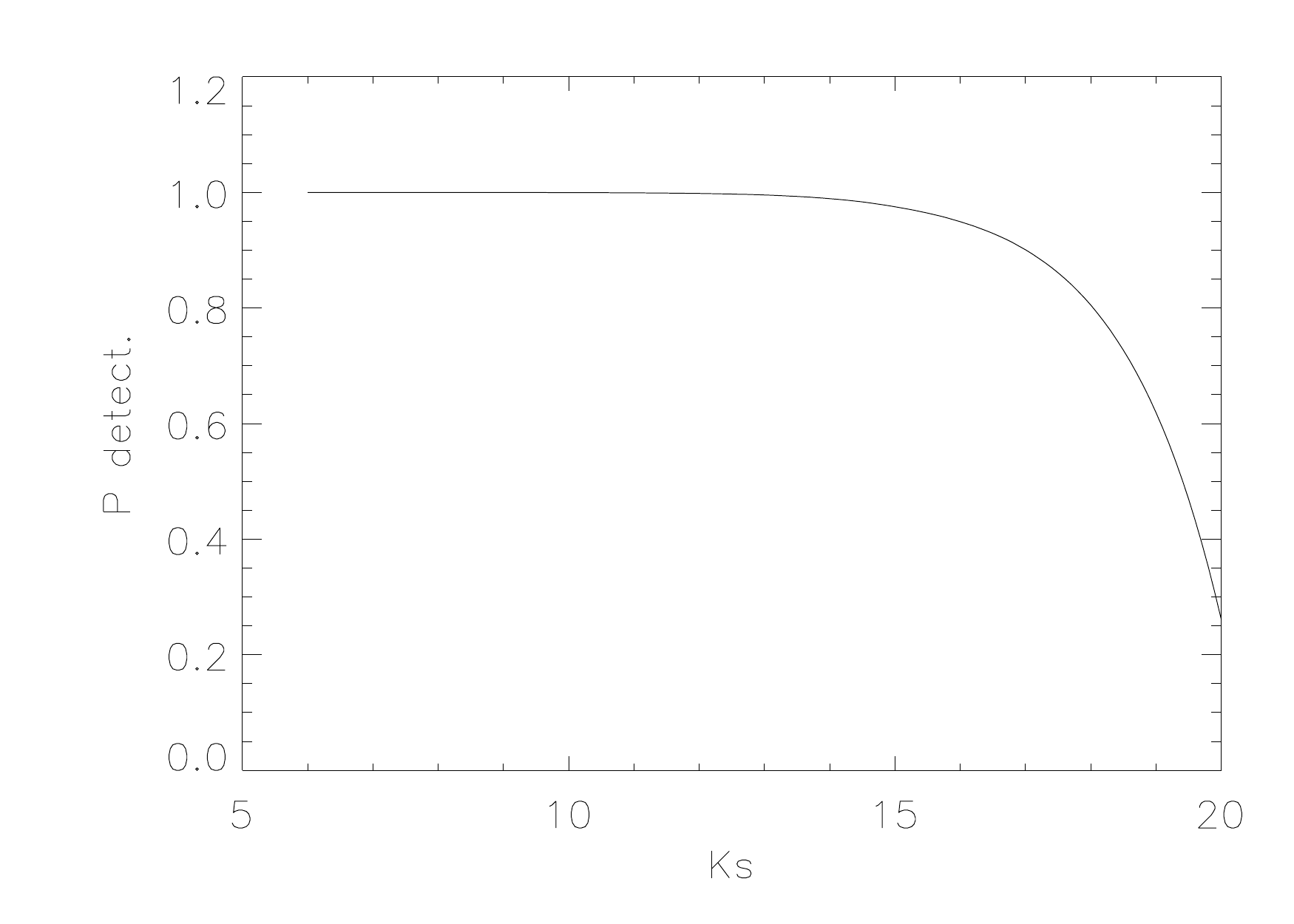}
   \caption{Detection probability as a function of the stellar magnitude resulting from the effect of the crowding in the Tr~14 fov.}
   \label{fig: pdetect}
   \end{figure}

\section{Observational biases}\label{sect: bias}
This section first describes an artificial star experiment 
designed to quantify the detection biases in the vicinity of the bright stars.
It also presents two simple models that generalise the results of the 
artificial star experiment and provide an estimate of the impact of
various observational biases on the results of this paper.
  
\subsection{Artificial star experiment}\label{sect: artif}

Most of the difference in the companion properties of massive and
lower-mass stars are found for companions in the range $13<K_\mathrm{S, sec}<14$.
In this section, we describe and analyse the results of an artificial 
star experiment that aims at better understanding the limitation of our
data in that range, justifying some of the choices made
in the previous section. 
The PSF of the brightest sources indeed show strong and extended wings,
which decrease the detection likelihood in the 
neighbourhood of a bright star. As a by-product, the results of this 
experiment also provide an
independent estimate of the photometric errors and of the completeness of
our catalogue for the tested parameter range, but this is not our main purpose.

The artificial star experiment follows a procedure similar to that presented in
\citet{MHS02, MOB08}.  In particular, stars with  known $H$ and $K_\mathrm{S}$
magnitudes  were simulated into the individual $H$ and $K_\mathrm{S}$ images using 
the  PSF of each  image and taking into account the quadratic 
dependence of the PSF with the position in each frame. 
 The entire  reduction  procedure  was then repeated
and  the artificial  stars  were  reduced  as  described in Sect.~\ref{sect: obs}.

To the first order, stars brighter than $K_\mathrm{S}=11$ in Tr~14 have masses of 10~\msun\ or more
(see e.g.\ Fig.~\ref{fig: cmd}). As in Sect.~\ref{sect: comp}, we adopted this limit for our massive star sample. 
We thus selected 14 bright stars with $9\le K_\mathrm{S}\le 11$. Around each of them we simulated 50 companions 
with $13\le K_\mathrm{S, in}\le 14$ and  $H_\mathrm{in}-K_\mathrm{S, in}=0.35$, spread in a 5\arcsec\ radius. 
The colour of the artificial stars were chosen to reproduce the colour of typical PMS stars in Tr~14, which are
the dominant type of sources in that magnitude range.

 Seven hundred artificial stars  were thus simulated in the $H$ and $K_\mathrm{S}$ images.
To optimise the computation time of this complicated procedure, all the artificial stars were added simultaneously. While this led to some heavier crowding than in the original field, this will be
taken into account in the analysis. This further allows us 
to study the detection biases using artificially controlled pairs.

\subsubsection{Separation distribution}\label{sect: artif_sep}
Figure~\ref{fig: artif_dmin} shows the cumulative distribution of the separations of the artificial 
stars with respect to one another, with respect to the closest source in the field and with 
respect to the closest bright source in the field. It shows that we are able to investigate 
various ranges of separation, from the crowding of the artificial 
stars among themselves to the effect of field density. 
We first focus on the closest detections. The main results of the artificial star experiment in this
respect are :
\begin{enumerate}

\item[-] The closest recovered pair of artificial stars is separated by 
$d_\mathrm{sep}=0.2$\arcsec, in good agreement with the IQ derived earlier,

\item[-] The recovery fraction of artificial pairs for which both components are further away
than 0.6\arcsec\ from any source in the image is better than 0.99 for $d_\mathrm{sep}>0.24$\arcsec,

\item[-] Excluding all  pairs of artificial stars with $d_\mathrm{sep}<0.3$\arcsec, the closest separation between a recovered artificial stars and a star at least as bright is 0.23\arcsec,

\item[-] Similarly,  the closest separation between a recovered artificial star and a massive star ($K_\mathrm{S}<11$) is 0.38\arcsec, in perfect agreement with what we found in our data (Sect.~\ref{sect: comp_chance}),

\item[-] All in all, there is a good agreement between the results of the artificial star experiment and the maximum reachable contrast 
as a function of the separation described empirically by Eq.~\ref{eq: contrast}.
\end{enumerate}
As a first result, the artificial star experiment allows us to validate up to $\Delta K_\mathrm{S}=5$~mag at least the empirical contrast {\it vs.} separation function introduced earlier. This result will be used below to build a first-order analytical model of the observational biases affecting our data.

\subsubsection{Recovery fraction}\label{sect: artif_recov}

To estimate the recovery fraction of solar-mass PMS stars in the wings of the brighter massive stars, we first
excluded all the close artificial pairs from our analysis. We also exclude all the artificial stars that fall closer than 0.3\arcsec\ from any source fainter than $K_\mathrm{S}=11$ in our data. With this we eliminate the uncertainties due to (i) crowding among the artificial stars and (ii) confusion between the artificial stars and the numerous field sources in our data. Almost 500 artificial stars are left, providing a decent coverage of the parameter space. 

Figure~\ref{fig: recov} shows the recovery fraction as a function of the separation to the closest bright star.  As mentioned earlier, the first detection occurs for a separation of 0.38\arcsec. For $d_\mathrm{sep}>0.53$\arcsec, all the artificial stars are recovered. Between 0.38\arcsec\ and 0.53\arcsec, the recovery fraction is approximately 0.4 and remains constant over the interval.

\subsubsection{Photometric uncertainties}\label{sect: artif_photom}

The  comparison between the artificial star input ($K_\mathrm{S, in}$) and recovered ($K_\mathrm{S, out}$) magnitudes provides us with a more realistic estimate of the  photometric errors (Fig.~\ref{fig: artif}). It reveals that the accuracy of both the retrieved magnitudes and of the colour term is significantly affected within $\approx 1$\arcsec\ from a bright star, the fainter companion being up to 0.1~mag too red. It also reveals a slight systematic shift in the retrieved colour and magnitudes, even at larger distance. Rejecting the few significantly deviant points resulting from the crowding, we obtained for the artificial stars more distant than 1.5\arcsec\ from the closest bright star
\begin{eqnarray}
H_\mathrm{out}-H_\mathrm{in}&=&0.008\pm 0.020 \\
K_\mathrm{S, out}-K_\mathrm{S, in}&=&0.008\pm 0.026\\
(H-K_\mathrm{S})_\mathrm{out}-(H-K_\mathrm{S})_\mathrm{in}&=&-0.021\pm 0.019,
\end{eqnarray}
Those are slightly larger deviations that the formal errors of the PSF photometry (Fig. ~\ref{fig: error}). The systematic increase of the $H-K_\mathrm{S}$ colour  when getting closer to the bright companion could result from a slightly better IQ obtained in the $K_\mathrm{S}$ band compared to the $H$ band, and thus a better subtraction of the bright star wings for a given separation.

\subsection{Analytical models}

\subsubsection{Impact of crowding}\label{sect: bias_crowd}

Given the very steep transition in the detection probability once the separation to a brighter source increases, and because the artificial star experiment generally agrees with the empirical detection limit given by Eq.~\ref{eq: dmin}, one can develop a very simple model to estimate the impact of the crowding in the field. It relies on the following hypotheses :
\begin{enumerate}
\item[-] A brighter star is always detected if falling on top of or very close to a fainter one,
\item[-] Non-detection is the consequence of shadowing by brighter stars in the fov,
\item[-] The spatial distribution of the stars in the field is random.
\end{enumerate} 
The detection probability of a source of a given magnitude $K_\mathrm{S}^0$ can then be written as 
\begin{equation}
P_\mathrm{detect.}(K_\mathrm{S}^0)=1-\frac{\Sigma_{K_\mathrm{S}\le K_\mathrm{S}^0}\left( \pi d^2(\Delta K_\mathrm{S})\right)}{A_\mathrm{fov}},
\label{eq: pdetect}
\end{equation}
where $d(\Delta K_\mathrm{S})$ is given by Eq.~\ref{eq: dmin} and where $A_\mathrm{fov}$ is the area of
the considered fov. In Eq.~\ref{eq: pdetect}, the sum is performed over all sources brighter than $K_\mathrm{S}^0$. Figure~\ref{fig: pdetect} shows the resulting detection probability. According to this model, completeness level of 0.99, 0.95 and 0.90 are reached for stars brighter than $K_\mathrm{S}=14$, 16 and 17 respectively. While our model is certainly too crude to adequately describe the faint end, it still shows that the crowding has a very limited impact in Tr~14. In particular, this indicates that the star counts used to fit the Tr~14 profile and to estimate the cluster core mass are unlikely to be significantly biased.

   \begin{figure}
   \centering
   \includegraphics[width=\columnwidth]{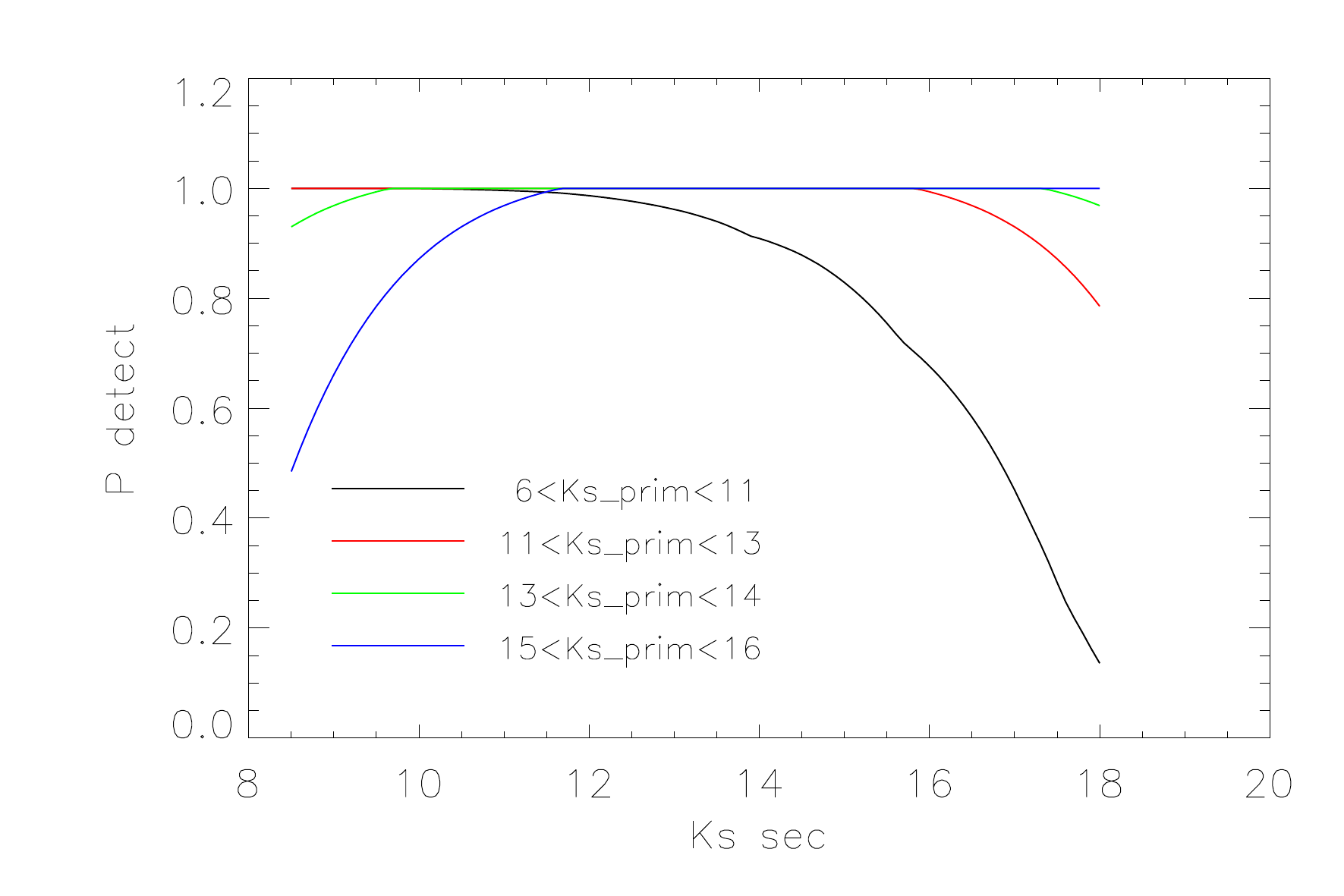}
   \caption{Detection probability model for pairs with separation in the range of 0.5\arcsec-2.5\arcsec\ and for various brightness ranges of the central star.}
   \label{fig: comp_bias}
   \end{figure}

\subsubsection{Companion detection threshold}\label{sect: bias_comp}

In Sect.~\ref{sect: comp_freq} to  \ref{sect: comp_spatial}, we focused our analysis to the 0.5-2.5\arcsec\ separations. In this section, we develop again a simple model to better quantify the impact of the observational biases in that range.
 As above, we use Eq.~\ref{eq: dmin} to define the region where a star outshines close fainter neighbours given the magnitude contrast of the pair. 
 In particular, the detection probability of a neighbour of magnitude $K_\mathrm{S, sec}$ in the vicinity of a $K_\mathrm{S, prim}$ central star can be modelled as the ratio between the area where one of the components does not outshine the other and the total area considered. For an annulus region $R_\mathrm{min}$--$R_\mathrm{max}$ around the central star, the detection probability can thus be written as
\begin{equation}
P_\mathrm{detect.}= \left\{
    \begin{array}{ll}
    1.0,          & \mathrm{if\ } d < R_\mathrm{min}''; \\ 
1.0- \frac{d^2(\Delta K_\mathrm{S})-R_\mathrm{min}^2} {R_\mathrm{max}^2-R_\mathrm{min}^2}  & \mathrm{if\ }  R_\mathrm{min} \le d < R_\mathrm{max}''; \\
    0.0,          & \mathrm{if\ } d \ge R_\mathrm{max}.
 \end{array} \right.
\label{eq: pdetect2}
\end{equation}
Figure~\ref{fig: comp_bias} compares the detection probability in the 0.5-2.5\arcsec\ separation range for the various central star-brightness ranges used earlier. For the bright primary interval, the curve displayed results from the average of $P_\mathrm{detect.}$ computed individually for all the bright primaries considered to build Figs.~\ref{fig: comp} and \ref{fig: mag_dist}. The central magnitude of the  primary intervals has been used for the other three categories.

Our results show that the detection of the companions to intermediate-mass stars ($11<K_\mathrm{S}<13$) and to solar-mass PMS stars ($13<K_\mathrm{S}<14$) stars is mostly unaffected. As expected, the lower mass PMS ($15<K_\mathrm{S, prim}<16$) are less likely to be found close to a bright stars. However, taking into account the number of bright stars in the field, one can prove this to be completely negligible. The largest biases are affecting the companions of bright stars.  $P_\mathrm{detect.}$ is passing below 0.9 at $K_\mathrm{S}=14$ and below 0.5 at $K_\mathrm{S}\approx17$.  One can further show that up to one companion per star is likely lost  at $K_\mathrm{S, sec}<16$ in Fig.~\ref{fig: comp}.

   \begin{figure}
   \centering
   \includegraphics[width=\columnwidth]{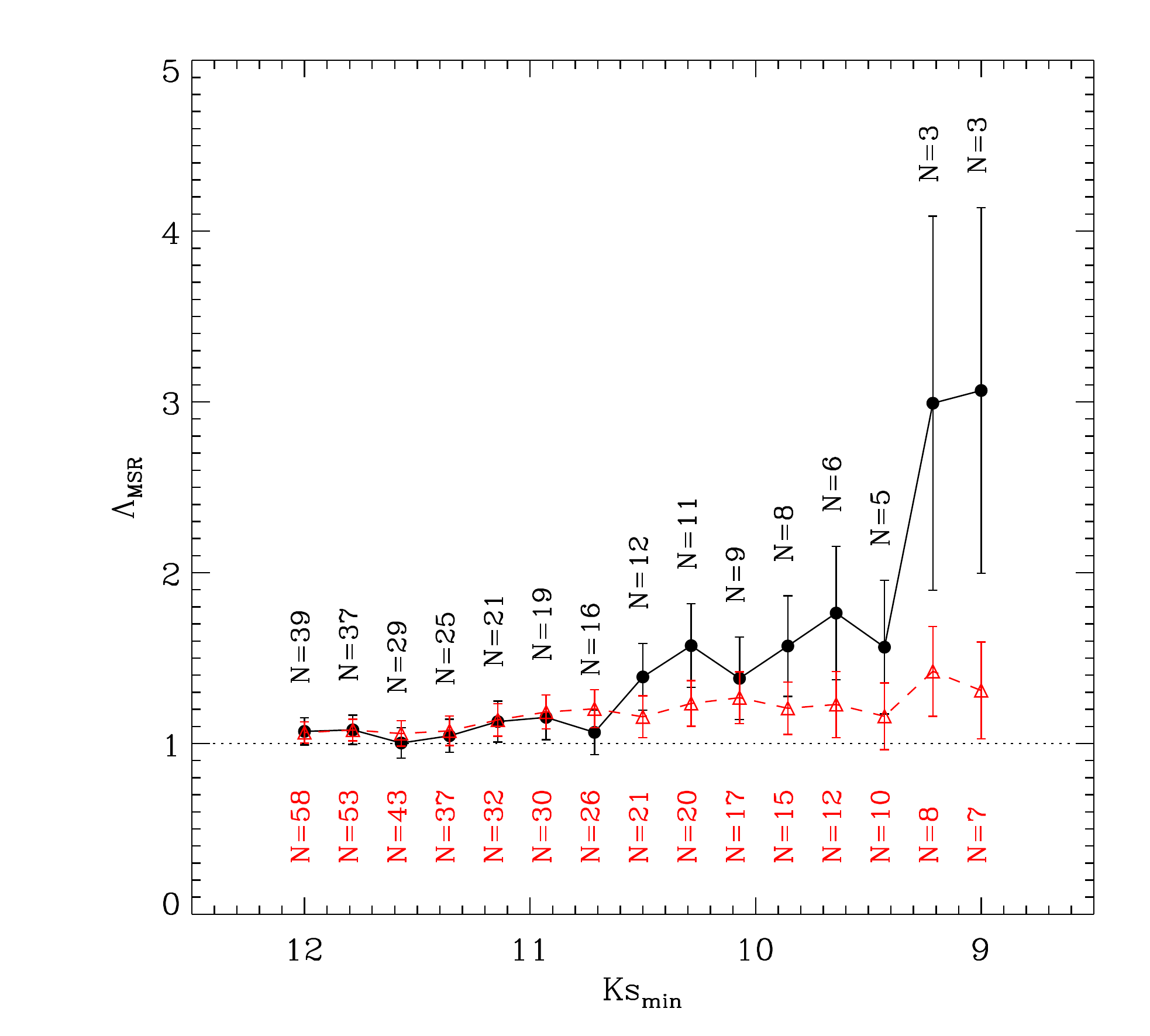}
   \caption{Evolution of $\Lambda_\mathrm{MSR}$ with the adopted magnitude limit $K_\mathrm{S, min}$ for the massive star sample. The circles and triangles indicate the results obtained with and without the colour and magnitude selection criteria of Fig.~\ref{fig: sigclip}. The upper and lower $N$ values give the number of stars considered in each step and for both samples. The dotted line indicates $\Lambda_\mathrm{MSR}=1$, i.e., no mass segregation.}
   \label{fig: mst}
   \end{figure}


\section{Mass segregation}\label{sect: mst}
As introduced in Sect.~\ref{ssect: eff}, the more massive MS stars seem more concentrated towards the cluster centre than the lower mass PMS stars. The best-fit EFF87 profiles (Table~\ref{tab: eff}) confirm that this difference is indeed significant at the 4$\sigma$ level. This could be interpreted as a hint for mass segregation, although \citet{AAL09} warned against hasty conclusions because numerous observational biases are actually favouring the detection of mass segregation, even in non-segregated clusters. 

\citet{AGP09_mst} recently introduced an alternative method to investigate mass segregation, which is insensitive to biases like the exact location of the cluster centre, and less sensitive (although quantification is still lacking) to the incompleteness effects. Their method compared the minimum spanning tree (MST), the shortest open path connecting all points of a sample, of the massive stars to the equivalent path of low mass stars (see \citeauthor{AGP09_mst} for a full description of the algorithm). The mass segregation ratio ($\Lambda_\mathrm{MSR}$), i.e. the ratio between the average random path length and that of the massive stars, allows them to quantify the deviation between the massive star sample and the reference sample. Following their approach, Fig.~\ref{fig: mst} displays the evolution of $\Lambda_\mathrm{MSR}$ with the adopted magnitude limit ($K_\mathrm{S, min}$) for the massive star sample. The reference distribution consists of stars with $14<K_\mathrm{S}<16$ and was drawn 500 times from our catalogue, with each sample containing the same number of stars as found in the bright sample. The dispersion obtained gives us the error bars on  $\Lambda_\mathrm{MSR}$ as displayed in Fig.~\ref{fig: mst}.  In the above procedure, we deliberately remained far from the limiting magnitude of our catalogue to minimize the completion biases. The method is in principle still affected by crowding and by the shadowing in the vicinity of bright stars. We showed  in Sect.~\ref{sect: artif_recov} however that the former effect had a very limited impact down to $K_\mathrm{S}<16$ at least. The effect of the shadowing  in the vicinity of bright stars is more difficult to estimate, although one can expect that the absolute number of $14<K_\mathrm{S}<16$ stars lost is proportionally very small compared to the number of stars in that interval. This results from the low number of bright stars and from the limited radius at which they can outshine a fainter neighbour.

We computed the MST first with and without the colour selection defined in Sect.~\ref{sect: clean} (Fig.~\ref{fig: mst}). Focussing on the most probable members (thus applying the colour selection), we found some indication of mass segregation down to $K_\mathrm{S, min}\approx10.5$~mag at the 1.5$\sigma$ level. As expected, the degree of mass segregation is increasing with the average brightness of a sample, thus with the mean stellar mass.
The largest mass segregation ratio is obtained for the few brightest, most massive stars after applying the membership selection. Yet the 1.5$\sigma$ confidence of this result remains at the limit of the detection.

Because the sensitivity of the MST to the completeness of a sample is not fully understood \citep{AGP09_mst}, we cannot draw firm conclusions. We note  however that two independent methods, profile fitting and MST analysis, both point towards mass segregation, as can be expected for the most massive stars in such a cluster. 

Obviously our observations only allow us to investigate the current mass segregation status of the cluster and we cannot distinguish whether this segregation, if confirmed, is primordial or is the product of early dynamical evolution. Given the expected cluster mass and size and its stellar contents as obtained in Sect.~\ref{ssect: eff} after colour selection, we estimated the typical dynamical friction time-scale $t_\mathrm{df}$ of 10~\msun\ and 20~\msun\ stars (corresponding to resp.\ $K_\mathrm{S}\approx11$~mag and 9.5~mag in our data). Following \citet{SpH71} and \citet{PZM02} , we obtained $t_\mathrm{df}\approx7.2\times10^5$~yr and $3.6\times10^5$~yr respectively. Considering the estimated age of the cluster, 3-5$\times 10^5$~yr, the dynamical friction time-scales agree with the results of Fig.~\ref{fig: mst}, where mass segregation begins to appear somewhere between 10~\msun and 20~\msun. As a consequence, if mass segregation is confirmed, it does not need to be primordial but can probably be explained by dynamical evolution.


\section{Summary and conclusions} \label{sect: ccl}

Using the ESO MCAO demonstrator MAD, we have acquired deep $H$ and $K_\mathrm{S}$ photometry of a 2\arcmin\ region around the central part of Tr~14. The average IQ of our campaign is about 0.2\arcsec\ and the dynamic range is about 10~mag. The image presented in Fig.~\ref{fig: fov} is by far the largest AO-corrected mosaic ever acquired. 

Using PSF photometry, we investigated the sensitivity of faint companions detected in the vicinity of bright sources. We derived several empirical relations that can be used as input for instrumental simulations, to estimate the performance of AO techniques versus seeing-limited techniques or, as done later in this paper, to build first-order analytical models of the impact of some observational biases. In particular, the contrast vs.\ separation limit has been validated over a 5 magnitude range by an artificial star experiment.

Despite a probably significant contamination by field stars, the Tr~14 CMD shows a very clear PMS population. Its location in the CMD can be reproduced by PMS isochrones with contraction ages of 3 to $5\times10^5$~yr. Interestingly, Tr~14 cannot be significantly further away than the distance obtained by \citet{CRV04} i.e., 2.5~kpc, as this would result in an even earlier contraction age. We derive the surface density profile of the cluster core and of different subpopulations. For stars brighter than $K_\mathrm{S}=18$~ mag, the surface density profiles are well reproduced by  EFF87 profiles over our full fov, and we provide quantitative constraints on the spatial extent of the cluster and on its stellar contents. Adopting the core-halo description suggested by \citet{AAV07}, we report that the transition between the core and the halo is not covered by our data, implying that the core is strictly dominating the density profile in a radius of 0.9~pc at least. Using colour criteria to select the most likely cluster members, the density profiles of the  more massive MS stars are best described by a power-law (or, equivalently, by an EFF87 profile with a very small core radius).

We also investigated the companionship properties in Tr~14. We showed that the number of companions and the pair association process is on average well reproduced by chance alignment from a uniform population randomly distributed across the field. Only stars with a brightness ratio close to unity or with a separation of less less than 0.5\arcsec\ cannot be explained by spurious alignment and are thus true binary candidates. This does not imply that large light-ratio and/or wider pairs do not exist, but rather that they cannot be individually disentangled with statistical arguments.  Still, 19\%\ of our massive star sample have a high probability physical companion. 

Focusing on the 0.5\arcsec-2.5\arcsec\ separation range, where the observational biases are  unable to invalidate our results, we compared the companion distributions of massive stars with those of lower mass stars. In Tr~14, the high-mass stars ($M>10$~\msun) tend to have more solar-mass companions than lower-mass comparison samples. Those companions are brighter on average, thus more massive. Finally, no difference could be found in the spatial distribution of the companions of low and high-mass stars.

Lastly, we employed the MST technique of \citet{AGP09_mst} to investigate possible mass segregation in Tr~14.  Again we found marginally significant results (at the 1.5$\sigma$ level), suggesting some degree of mass segregation for the more massive stars of the cluster ($M>10$~\msun) . Although the sensitivity of the method to incompleteness is still not fully quantified, we note that early dynamical evolution can reproduce the observed hints of mass segregation in Tr~14, despite the cluster's young age.

\begin{acknowledgements}     
The authors are greatly indebted to Paola Amico and to the MAD SD team for their excellent support during the preparation and execution of the observations.  We also express our thanks to the referee for his help, which clarified the manuscript, and to Dr. Joana Ascenso for useful discussions.
\end{acknowledgements}

\end{document}